\documentclass[reprint,10pt,aps,twocolumn,nofootinbib,superscriptaddress,showkeys]{revtex4-1}

\usepackage{newtxtext}
\usepackage[slantedGreek]{newtxmath}
\usepackage[nodayofweek]{datetime}
\newdateformat{myDate}{\THEDAY\ \monthname[\THEMONTH] \THEYEAR} %

\usepackage{amssymb,amsmath}
\usepackage[colorlinks,citecolor=blue,urlcolor=black,linkcolor=blue,breaklinks]{hyperref}
\hypersetup{pdftex,colorlinks,breaklinks,
            pdftitle={Nonlinear feature filtering (NPHIL)},
            pdfauthor={C. Poelking},
            pdfsubject={},
            pdfkeywords={},
            urlcolor=[rgb]{0,0,0},
            citecolor=[rgb]{0.0,0.4,1.0},
            linkcolor=[rgb]{0.0,0.4,1.0}}
\usepackage{graphicx}
\usepackage{bm}
\usepackage{units}
\usepackage{dsfont}
\usepackage{lineno}
\newcommand{\pyield}{\varepsilon}
\newcommand{\denseitem}{\vspace{-1pt}\item}
\newcommand{\dx}{\mathrm{d}}
\newcommand{\inserteq}[1]{
 \begin{linenomath*}
 \begin{align}
  #1
 \end{align}
 \end{linenomath*}
}

\newcommand{\sifigdeconv}{1}
\newcommand{\sifigstats}{2}
\newcommand{\sifigsyn}{3}
\newcommand{\sifigcv}{5}

\begin{document}

\title{\LARGE Noisy, sparse, nonlinear: Navigating the Bermuda \\
Triangle of physical inference with deep filtering}
\author{Carl Poelking}
\email{Please address correspondence to cp605@cam.ac.uk}
\affiliation{Department of Chemistry, University of Cambridge, UK}
%\affiliation{Department of Chemistry, University of Cambridge, Lensfield Road, CB2 1EW Cambridge (UK)}
\author{Yehia Amar} 
\author{Alexei Lapkin}
\affiliation{Department of Chemical Engineering and Biotechnology, University of Cambridge, UK}
%\affiliation{Department of Chemical Engineering and Biotechnology, University of Cambridge, Philippa Fawcett Drive, CB3 0AS Cambridge (UK)}
\author{Lucy Colwell}
\affiliation{Department of Chemistry, University of Cambridge, UK}
%\affiliation{Department of Chemistry, University of Cambridge, Lensfield Road, CB2 1EW Cambridge (UK)}
\affiliation{Google Research, Mountain View, CA}

\date{\myDate\today}
\begin{abstract}
Capturing the microscopic interactions that determine molecular reactivity poses a challenge across the physical sciences. Even a basic understanding of the underlying reaction mechanisms can substantially accelerate materials and compound design, including the development of new catalysts or drugs. Given the difficulties routinely faced by both experimental and theoretical investigations that aim to improve our mechanistic understanding of a reaction, recent advances have focused on data-driven routes to derive structure-property relationships directly from high-throughput screens. However, even these high-quality, high-volume data are noisy, ulteriorly sparse and biased -- placing them in a regime where machine-learning is extremely challenging. Here we show that a statistical approach based on deep filtering of nonlinear feature networks results in physicochemical models that are more robust, transparent and generalize better than standard machine-learning architectures. Using diligent descriptor design and data post-processing, we exemplify the approach using both literature and fresh data on asymmetric catalytic hydrogenation, Palladium-catalyzed cross-coupling reactions, and drug-drug synergy. We illustrate how the sparse models uncovered by the filtering help us formulate physicochemical reaction ``pharmacophores'', investigate experimental bias and derive strategies for mechanism detection and classification.
\end{abstract}

\maketitle

\section{Introduction}
\label{sec:introduction}

Applications of machine-learning to complex materials and reaction systems have recently drawn intense interest from both basic and industrial research, with substantial progress being made in computational method development and integration with high-throughput synthesis and analytics~\cite{isbrandt_high_2019,granda_controlling_2018,gromski_how_2019,allen_power_2019,zahrt_prediction_2019}. Data-driven strategies enable in-silico predictions in reactive multi-component environments (such as palladium-catalyzed cross-coupling reactions or enantioselective synthesis) that can elude direct quantum-mechanical or atomistic approaches~\cite{reid_comparing_2018,rosales_rapid_2019}, with application scenarios ranging from computer-assisted retrosynthesis~\cite{segler_planning_2018} and the optimization of reaction conditions~\cite{zhou_optimizing_2017} to in-silico drug~\cite{chen_rise_2018,ekins_exploiting_2019,bartok_machine_2017}, catalyst~\cite{goldsmith_machine_2018,meyer_machine_2018} and materials~\cite{raccuglia_machine-learning-assisted_2016} discovery. Even though application objectives are diverse, the computational architectures have in common that they rely on trained models that predict reaction products~\cite{coley_graph-convolutional_2019} and performance~\cite{ahneman_predicting_2018} given an appropriately chosen or learned representation of a set of molecular compounds~\cite{sigman_development_2016,gomez-bombarelli_automatic_2018}.

% NOISE AND NONLINEARITY
Required to be accurate, robust and transferable, these machine-learned models of molecular reactivity are faced with a number of challenges~\cite{sigman_development_2016}. One such challenge lies in the strongly nonlinear behaviour that reaction networks often display due to the complex interactions that determine rate-limiting steps, including potential activity cliffs that occur as one rate-limiting step gives way to another. The nature of these reaction bottlenecks implies that the best solution functional, expressed in terms of the molecular descriptors, can be extremely sparse -- i.e., it may depend on only a small subset of the large set of properties that capture molecular behaviour~\cite{harper_predicting_2011,milo_data-intensive_2015}. Uncovering such sparse solutions is, however, thwarted by low signal-to-noise ratios -- an unfortunate but common side effect as experimental accuracy is balanced against the need for high throughput. The resulting spurious patterns are not only confusing to the model, but also the modeller: As the nonlinear distance (measured, e.g., in terms of polynomial power) between input variables and output property increases, the accuracy of the machine-learning model tends to react in exactly the same way as when facing a decreasing signal-to-noise ratio: The performance margin over a random null model decreases, leaving us wondering whether the poor performance is limited by dataset noise (and hence outside our immediate control), or by the model's capacity to find an appropriate parametrization.

\begin{figure*}[t]
\centering
\includegraphics[width=1.0\linewidth]{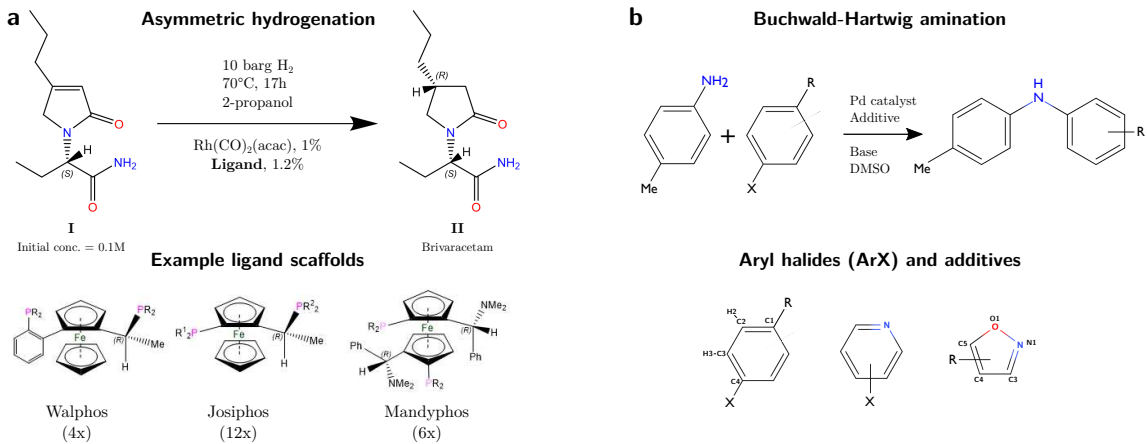}
\caption{ (a) Simplified reaction scheme and conditions for the asymmetric hydrogenation of a Brivaracetam precursor, with example molecular scaffolds of three ligand families: Walphos, Josiphos, Mandyphos. (b) Reaction schematic for the Buchwald-Hartwig cross-coupling amination reaction. }
\label{fig:fig_reactions}
\end{figure*}

But it can get worse: Whereas nonlinearity and noise tend to impact model performance in a random, unsystematic way, data and model bias, by contrast, systematically steer the model away from the ``correct'' solution -- with potentially serious implications when machine learning is employed in medical or social contexts~\cite{challen_artificial_2019,courtland_bias_2018}. In a molecular context, methodical studies of bias artifacts are few and far between~\cite{wallach_most_2018,mccloskey_using_2018}. Here we limit the discussion to the following three types:
\begin{enumerate}
 \denseitem {\it Sampling bias} is a {\it dataset artifact} that affects all finite-sized datasets: It tends to deteriorate a model's capacity to generalize because it learns patterns inadvertently or unknowingly introduced into the dataset due to suboptimal sample selection. 
 \denseitem {\it Structural bias} is a {\it modelling artifact} arising in the context of many materials datasets: It results whenever the abstraction from (atomistic) structure to property (and, finally, to target) fails: Or, in other words, when predictions are based on motif, scaffold or warhead recognition rather than the physical properties associated with those motifs.
 \denseitem {\it Combination bias} is a {\it modelling artifact} that arises in multi-component reaction systems consisting of more than one variable factor (e.g., a base, a solvent, a catalyst and a ligand). We say that a model is combinatorially biased if it predicts the outcome of a new molecular combination based on {\em those} of its components that have previously been trained on in an identical or similar sub-configuration.
\end{enumerate}
Whatever the source, bias reduces a model's capability to generalize, while artifically inflating performance metrics -- especially if inappropriate cross-validation techniques are used. Even though certain sources of bias (such as sampling bias) can only be ruled out in the big-data limit, there are measures one can take to avoid bias when first collecting the data, as well as to decrease the susceptibility to bias during the modelling stage. The risk of sampling bias, for example, is routinely reduced during data collection by ensuring sample diversity -- i.e., maximising the number of distinct scaffolds and substituents while also taking into account physicochemical diversity. Structural bias, on the other hand, is often a consequence of applying high-dimensional structural descriptors to sparse data. Even physicochemical features (in particular, NMR, IR, or electrostatic fingerprinting) can be ``misused'' by the model as a proxy for structural descriptors -- resulting in a ``regression to structure'' that defeats the original purpose of the physicochemical approach.

% THIS WORK
In this work, we present a learning framework derived from nonlinear feature networks that we believe addresses many of these challenges (nonlinearity, noise and bias) -- resulting in a machine-learning architecture that is simple, interpretable and yet generalizes well across compound classes. Making use of bias-reducing data post-processing techniques and careful descriptor design, the model performs nonparametric statistical filtering to detect extreme covariance events within a deep feature network when offset against an explicitly simulated null background. Designed to capture physically relevant patterns, these extreme events indicate sparse functions that can subsequently inform regression and classification frameworks. Applications covering cross-coupling amination reactions, asymmetric catalysis and drug-drug synergy illustrate the excellent performance even when subjected to challenging cross-validation rules. Inspection and visualization of the feature networks meanwhile allows us to derive physicochemical reaction ``pharmacophores'' to glean deeper insight into molecular reactivity.

\section{Results and Analysis}

\subsection{Reaction systems}

We model molecular (re)activity in three different settings, targeting (a) the diastereoselectivity of a rhodium-catalyzed asymmetric hydrogenation,  (b) the reaction yield of a palladium-catalyzed Buchwald-Hartwig cross-coupling amination, and (c) the growth-inhibitory synergy of anti-cancer drug combinations. In all cases, the reaction inputs are compound combinations, with the variable molecular factors determined by the respective application and synthesis objectives.

First, the rhodium-catalyzed asymmetric hydrogenation involves the hydrogenation of a chiral $\alpha$-$\beta$ unsaturated $\gamma$-lactam (I) (see Fig.~\ref{fig:fig_reactions}a) used to produce UCB Pharma's anti-epileptic drug Brivaracetam (II)~\cite{stephen_brivaracetam:_2018}. We have recently reported an operando NMR study showing the effectiveness of the Rh(CO)2(acac) precursor for asymmetric hydrogenation, and demonstrated the beneficial effect of alcohol solvents on diastereomeric excess and conversion~\cite{amar_machine_2019,amar_accelerating_2018}. However, yet superseeding the solvent effect, a bisphosphine ligand is the primary factor in driving reaction performance, with the binding modes and transition states of the catalyst-ligand-substrate complex identified as rate- and diastereo-determining. Predicting the effect of the bisphosphine ligand will therefore be the key focus of our modelling efforts.

Second, the Buchwald-Hartwig reaction considers molecular combinations consisting of an aryl halide (ArX) as substrate, an isoxazole additive, a ligand and a base (see Fig.~\ref{fig:fig_reactions}b). As detailed by Ahneman et al.~\cite{ahneman_predicting_2018}, this choice of variable inputs follows a Glorius fragment additive screening approach designed to model the detrimental effect of isoxazoles (or, more generally: five-membered heterocycles) on the yield of the amination. Incorporating the isoxazole as a distinct molecule instead of as a substrate moiety simplifies the experimental and synthetic procedure while covering a larger fraction of the relevant chemical space.

Third, the growth-inhibitory activity of anti-cancer drug combinations is explored using the NCI-ALMANAC (A Large Matrix of Anti-Neoplastic Agent Combinations), which tested more than 5000 combinations of FDA-approved cancer drugs across 60 human cell lines~\cite{holbeck_national_2017}. The assay quantifies whether two compounds are, as a combination, more or less effective than their calculated additive effect. Synergistic, additive or antagonistic action is detected experimentally by evaluating the growth-inhibitory effect using compound-compound concentration matrices (taking into account the concentration-dependent single-agent effects) based on a modified version of the Bliss-independence model. Instead of deriving cell-line specific models, our focus will be to derive generic rules that determine compound synergy by aggregating the data across all cell lines.

\subsubsection*{Yield deconvolution} Reactants or reagents with weak cross-coupling superseeding the effect of less important reagent classes is a key reason why multicomponent reaction systems can be less ``combinatorial'' than they are occasionally made out to be~\cite{allen_power_2019}. Not always easily noticed, this same superseeding effect renders machine-learning frameworks susceptible to combination bias, and a previously derived machine-learning model of the Buchwald-Hartwig system has as a result been critized for not performing better than a random control~\cite{note:chuang_comment_2018}. 

To guard against such bias, we employ a data post-processing step which partitions the target function (for the Buchwald-Hartwig system this would be the yield) onto unimolecular and bimolecular terms. This approach can be derived rigorously as a decomposition of an effective Gibbs free energy (see the SI appendix for details): For a generic three-factor system $(X_1,X_2,X_3)$ (where, e.g., $X_1$ indicates the substrate, $X_2$ the ligand, etc.), this decomposition when performed to second order reads
\inserteq{
 y(X_1 X_2 X_3) & \simeq y_0 + \pyield_1^{(1)}(X_1) + \pyield_1^{(2)}(X_2) + \pyield_1^{(3)}(X_3) + \nonumber \\
  + \pyield_{2}^{(1,2)}&(X_1 X_2) + \pyield_{2}^{(1,3)}(X_1 X_3) + \pyield_{2}^{(2,3)}(X_2 X_3).
}
Here $\pyield_1^{(i)}$ is the unimolecular yield function of compound species $i$, $\pyield_2^{(i,j)}$ the contribution due to the interaction (synergy) between two compounds of species $i$ and $j$; $y_0$ is the average yield observed across the dataset. 

Applied to the Buchwald-Hartwig system (with four instead of three factors), the deconvolution quantifies the relative importance of the four reactant classes regarding reaction outcome. The distribution of the partial yield terms (see Fig.~{\sifigdeconv} of the SI appendix) highlights that the aryl halide accounts for the largest variation in yield, followed by the additive. The deconvolution also illustrates that the entire yield is largely accounted for by only unimolecular terms: As these are averages over a large number of samples, they are expected to have a smaller statistical error than individual measurements. The accuracy in predicting these partial unimolecular yields instead of the total yield function is a significantly more robust performance measure that reduces susceptibility to bias hazards and measurement noise.

\begin{figure}[t]
\centering
\includegraphics[width=1.0\linewidth]{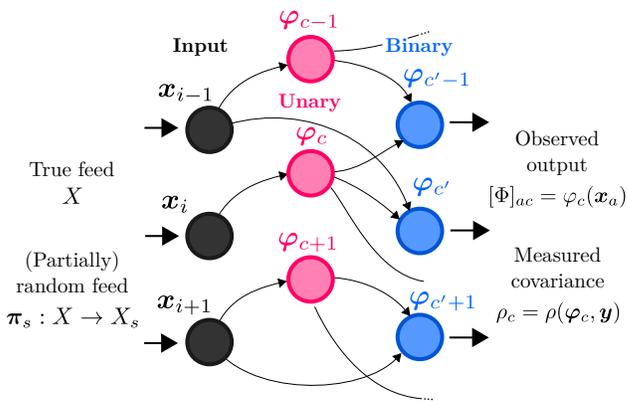}
\caption{Feature network schematic with one unary and one binary layer applied to a design matrix $X$ with targets $\bm{y}$; $i$ enumerates descriptor components, $c$ feature nodes, $a$ data samples. $\bm{\pi}_s$ is a vector of permutation operators. Combining true and partially randomized data feeds, the network output is finally submitted to the filtering procedure described in the main text.}
\label{fig:fig_network_schematic}
\end{figure}

\begin{figure*}[t]
\centering
\includegraphics[width=1.0\linewidth]{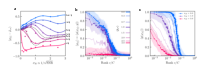}
\caption{Synthetic reaction data. (a) Impact of noise level $\sigma_R$ on the percentiles of the distribution of $(\rho_{[1]} - \tilde{\rho}_{c^*})$, which is the difference between the true covariance $\rho_{[1]}$ (covariance between the top-ranked feature and unperturbed target function $y$) and the sample covariance $\tilde{\rho}_{c^*}$ (covariance between the true feature $\varphi_{c^*}$ and perturbed target $\tilde{y}$). The solid lines are for sample sizes of $N=30$. Dashed lines indicate the median (50th percentile) associated with reduced sample sizes of $N=10,20$. (b) Median of the true covariance $\rho_{[c]}$ as a function of rank. Unlike $\langle \tilde{\rho} \rangle_{50}$, $\langle \rho \rangle_{50}$ experiences a sudden drop as the noise level rises above $\sigma_R = 1$. This breakdown is successfully picked up by (c) the median of the assigned confidences $\langle 1 - p(\varphi_{[c]}) \rangle_{50}$. Note that for panels e-f, the different solid curves correspond to noise levels $\sigma_R = 0.0$ (top, blue trend) to $\sigma_R = 3$ (bottom, red trend). For noise levels $\sigma_R = 0, 1, 3$ (dotted lines) confidence intervals of (for better distinguishability) half the standard deviation are shown.}
\label{fig:fig_stats}
\end{figure*}

\subsection{Feature network filtering}

The feature network filter addresses some of the key challenges that surround reactivity datasets. The three key coping strategies are as follows:
\begin{enumerate}
 \denseitem Data nonlinearity:  As a simple linear relationship is unlikely to capture reactive behaviour, a combinatorial feature network systematically generates increasingly complex nonlinear features/functions from a set of base variables/descriptors. These functions sample the space which we need to search for appropriate solutions.
 \denseitem Data sparsity: Sparse coverage of the relevant chemical space means that the sampling strength is generally insufficient to construct ``dense'' models such as parameter-heavy neural networks. Instead, we subsample and rank the functions generated by the feature network in order to obtain a sparse approximation to the target function and at the same time guard against structural bias.
 \denseitem Data noise: A low signal-to-noise ratio in connection with sparse data coverage causes spurious correlations that need to be identified. By resampling the correlations measured across the network and offsetting them against an explicitly simulated background (null) distribution, we assess which nodes of the network give rise to a physical signal, given the context of the {\it entire} network. This filtering ensures model viability even before the cross-validation stage.
\end{enumerate}

First, the construction of the feature network proceeds as follows: A graph incrementally generates nonlinear features $\varphi_{c}(x_1, \dots, x_d)$ from a set of $d$ input descriptors $x_{i}$ via a sequence of unary and binary operations (see Fig.~\ref{fig:fig_network_schematic}). The graph is constructed in a combinatorially exhaustive manner with certain restrictions regarding the allowed operations: The features should be scale-invariant (i.e., their covariance with the target should not dependent on the choice of units); they should be non-redundant, non-complex, finite and mathematically sound, as to be verified by symbolic algebra. The unary operations are $U \in \{\exp(*), \log(*), \sqrt{*}, (*)^{-1}, (*)^2, |*|\}$, binary operations include $B \in \{+,-,\times,\div\}$. We note that massive feature generation of this type has previously been applied to low-noise materials datasets in the context of structure and metal/insulator classification of (octet) binary crystals~\cite{ghiringhelli_big_2015, ouyang_sisso:_2018}.

The statistical analysis of the feature network is based on a covariance measure $\rho(\bm{\varphi}_c, \bm{y})$ that quantifies the sample covariance between an output function $\bm{\varphi}_c = (\varphi_{1,c}, \dots, \varphi_{N,c})^t$ of the network and target property $\bm{y} = (y_1, \dots, y_N)^t$ over $N$ training samples. Here we consider as covariance measure the Pearson correlation $\rho_p$ in the case of regression tasks (continuous $y$) and a signed AUC metric $\rho_\pm = (2 \mathrm{AUC}-1)$ for classification tasks (binary $y$). The filtering assumes that, for each function $\varphi_c$ of the network, the covariance $\rho_c$ between $\varphi_c(x_1, \dots, x_d)$ and $y$ can be decomposed such that
\inserteq{
 \rho_c & = \rho_{c,0} + \sum_i \rho_{c,i} + \sum_{i < j} \rho_{c,ij} + \nonumber \\
  &+ \sum_{i < j < k} \rho_{c,ijk} + \dots + \rho_{c, 1\dots d},
 \label{eq:rho_decomp}
} % end eq
where $\rho_{c,0}$ is a null covariance, and covariance terms $\rho_{c, i\dots j}$ quantify the joint covariance contribution due to higher-order correlations of the input descriptors $x_i, \dots, x_j$ with the target $y$. The null covariance highlights that even in the absence of any physical signal (where $\rho_{c, i\dots j} = 0$ by definition), $\rho_c$ follows a distribution that becomes light-tailed as the sample size decreases. This ``random-physics'' background probability density $p(\rho_{c,0})$ can formally be calculated as
\inserteq{
 p(\rho_{c,0}) = \int \dx x_1 \dots \dx x_d \delta[\rho_{c,0} - \rho(x_1, \dots, x_d)] \prod_{i=1}^d p(x_i). \nonumber
} % end eq
Permutation vectors $\bm{\pi} = (\pi_1, \dots, \pi_d)$ with $\pi_i \in \mathrm{S}_N$ are the natural way to estimate the marginal distributions $p(x_i)$ of input descriptor $x_i$ and hence the distribution of null covariances $p(\rho_{c,0})$: A large number of $S$ randomized instances $X_s$ of the data matrix $[X]_{ai} = x_{ai}$ ($1\leq a \leq N$ indexes data samples) are generated by permuting the entries along all columns $i$ independently according to a randomly sampled permutation $\pi_{s,i}$. For each random instance $s$ generated by $\bm{\pi}_s$, the feature graph is evaluated using the randomized inputs, and covariances $\rho_{s,c}$ are recorded for all output nodes $c$. By aggregating the covariance results from the $S$ random data feeds, we construct the order statistics $\rho_{[s],c}$ for each function, $(\rho_{1,c} \dots \rho_{S,c}) \rightarrow (\rho_{[1],c} \dots \rho_{[S],c})$,
% \inserteq{
% (\rho_{1,c} \dots \rho_{S,c}) \rightarrow (\rho_{[1],c} \dots \rho_{[S],c}),
% } % end eq
where $\rho_{[s],c}$ is the $s$'th-largest (by magnitude) covariance sampled for $\varphi_c$. The sample complementary distribution function underlying this sequence is approximated in the usual way as $\bar{F}_c(\rho=\rho_{[s],c}) = \frac{s}{S}$ (see Fig.~S{\sifigstats}a of the SI appendix).
%This distribution is shown in Fig.~\ref{fig:fig_stats}a for two example functions, together with the correlations $\rho^*_c$ measured for the true, physical data matrix $X$.

In order to derive probabilities for observing ``extreme'' events (correlations above a certain threshold which indicate physically meaningful features), a possible route onwards would be to fit the tail of $\bar{F}_c$ to a predefined functional form: Natural candidates are the regularized incomplete beta function, or the generalized Pareto distribution as used by the peak-over-threshold approach in extreme-value theory~\cite{coles_introduction_2001}. In practice, however, not all nodes can be appropriately described by these distributions. This may be due to, e.g., the limited or discrete value range of certain features. Furthermore, the tail may be dominated by permutations that (nearly or exactly) reproduce the true, physical state, giving rise to non-standard tail shapes and reducing the quality of the fit if the threshold is not chosen optimally. 

To overcome these difficulties and avoid assumptions regarding distribution type, we instead follow a non-parametric approach: As test statistic we define a tail exceedence measure (inspired by the peak-over-threshold approach from extreme-value theory):
\inserteq{
\varepsilon_{s,c} = \varepsilon_c(\bm{\pi}_s) = \frac{1}{r_t} \sum_{r=1}^{r_t} \frac{|\rho_{s,c}|-|\rho_{[r],c}|}{|\rho_{[r],c}|}.
} % end eq
The sum is over the tail region with cutoff $r_t \ll S$, here set implicitly via a threshold probability $p_t = \bar{F}_c(\rho_{[r_t]})$. In practice, $p_t \simeq 0.01$ (in which case the $1\%$ largest correlation observations are said to constitute the tail region for each channel) seems to produce robust confidence estimates.
\begin{figure*}[t]%[tbhp]

\centering
\includegraphics[width=1.0\linewidth]{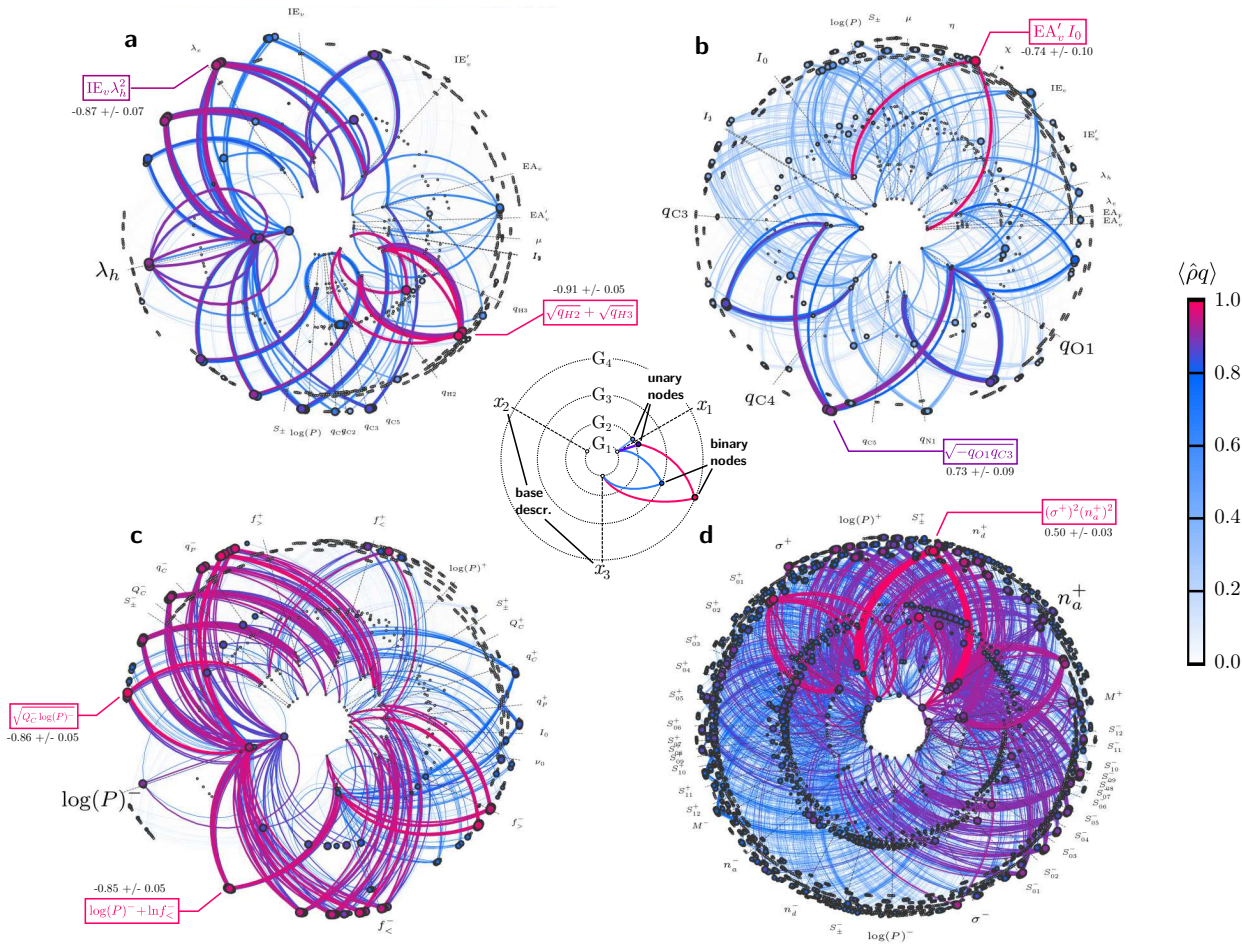}
\caption{ Feature-network visualizations for the (a) aryl halide and (b) additive partial yields, the (c) ligand diastereoselectivity and (d) anti-cancer synergy classification. Nodes are indicated with circles, links between nodes as arches. Root nodes are located on the innermost circle (as labelled by the dashed lines directed radially outwards), later generations on concentric circles of increasing radii. The size and colouring of a node $\varphi_c$ and the links connecting this node to its parents follow from its normalized signal strength $|\hat{\rho}_c q_c| = |\rho_c q_c / \rho_\mathrm{max}|$, where $\rho_\mathrm{max}$ is the largest-magnitude correlation sampled by the network. The angle subtended by the root nodes is proportional to the aggregated signal strength of all its descendants. Mathematical symbols are resolved in Table~\ref{tab:descriptors}. The numbers below labelled nodes are correlation values. }
\label{fig:fig_visual}
\end{figure*}

This exceedence measure is used as a global test statistic that normalizes the covariance output from the different nodes: Calculating only the null probability for observing $\rho_c^*$ given $\bar{F}_c(|\rho|)$ (i.e., the single-channel distribution) would not be sufficient to assess whether this correlation derives from a physical signal. Instead, the magnitude of $\rho_c^*$ needs to be evaluated in the context of the entire network. To do so, we collect the exceedences from all nodes $c$ to evaluate the order statistic for a random instance $s$: $(\varepsilon_{s,1} \dots \varepsilon_{s,c}) \rightarrow (\varepsilon_{s,[1]} \dots \varepsilon_{s,[c]})$.
% \inserteq{
% (\varepsilon_{s,1} \dots \varepsilon_{s,c}) \rightarrow (\varepsilon_{s,[1]} \dots \varepsilon_{s,[c]}).
% } % end eq
By aggregating this output across all $S$ random instances, we obtain the sample distribution functions of the ranked exceedences, $\bar{F}_{[c]}(\varepsilon) = \textrm{Prob}(\varepsilon_{[c]} \geq \varepsilon)$, for all ranks $c \leq C$ (see Fig.~S{\sifigstats}b of the SI appendix).
%\inserteq{
%\bar{F}_{[c]}(\varepsilon) = \textrm{Prob}(\varepsilon_{[c]} \geq \varepsilon).
%} % end eq
%These $C$ different distributions functions are represented in Fig.~\ref{fig:fig_stats}b via their mean (grey circles) and standard deviation (shaded area) together with the ranked exceedences $\varepsilon_{[c]}^*$ measured for the physical realization of the graph.

The output covariances measured for the true data matrix $X$ are finally evaluated against the sample exceedence distribution function of rank $[c] = [1]$ (shown in Fig.~S{\sifigstats}c). To this end, the {\it confidence} $q(\varphi_c) = 1 - \bar{F}_{[1]}(\varepsilon_c)$
%\inserteq{
%q(\varphi_c) = 1 - \bar{F}_{[1]}(\varepsilon_c), 
%} % end eq
quantifies our belief that the covariance observation for node $c$ reflects a physical signal in view of the large number of functions tested by the feature network. Note that we can in principle calculate a rank-specific confidence $q_{[r]}(\varphi_c)$ for the feature nodes, but it is at this point still unclear to which extent this rank-specific information could be usefully exploited.

In summary, the output of the feature network analysis is, first, a covariance observation $\rho^*_c$ with error $\Delta \rho^*_c$ (calculated via bootstrapping) for each function $\varphi_c$ and, second, a confidence $q_c = q(\varphi_c)$ with error $\Delta q_c$ that this feature constitutes a physical signal. Third, we can quantify the relative importance of the different base descriptors $x_i, \dots, x_j$ on which a function $\varphi_c$ depends by explicity constructing the decomposition in eq.~\ref{eq:rho_decomp} (see SI Appendix for details). When ranked according to a covariance measure $\rho$ (in line with the Sure-Independence-Screening approach developed by Fan {\it et al.}~\cite{fan_sure_2008}), this decomposition helps us discard functions where some of the base variables do not make any significant contribution to the measured signal.

\subsubsection*{Synthetic datasets} We briefly turn to synthetically generated data in order to gain insight into the performance and reliability of the technique. The datasets are generated using the following protocol: The descriptor consists of $d=10$ base variables, sampled such that five are strictly positive, $x \in (0,+t]$, and five positive or negative, $x \in [-0.5 t,0.5 t]$, where $t=10$ emphasises the difference between logarithmic, exponential and linear transformations. The base variables make up the input nodes of a feature network with one layer of unary and one layer of binary operators, resulting in a total of around $C=2000$ nodes. One of these nodes ($c^*$) is picked at random as the target function $y(\bm{x}) \propto \varphi_{c^*}(x_1 \dots x_d)$. This target function is z-scored and subsequently perturbed with Gaussian white noise of variance $\sigma_R^2$: $\tilde{y} = y + \mathcal{N}(0, \sigma_R^2)$. The width is varied in ten steps from $\sigma_R = 0$ to $3$. Three different dataset sizes are considered, $N=10, 20, 30$. For each parameter pair $(\sigma_R, N)$, 150 independent datasets are generated and analysed by the network, every time with a new generating function picked from among the nodes.

The results of the meta-analysis are summarized in Fig.~\ref{fig:fig_stats}a-c. The first quantity we investigate, $\rho_{[1]}-\tilde{\rho}_{c^*}$, measures the quality of the signal extracted from the dataset as the noise level increases: Here $\rho_{[1]}$ is the {\em true} covariance (calculated from a much larger sample) between the top-ranked feature and the {\em true} (i.e., unperturbed) generating function $y$. This implies that, if the true function happens to be ranked first, $\rho_{[1]} = 1$ holds independently of the noise level. From $\rho_{[1]}$ we subtract $\tilde{\rho}_{c^*}$, which is the sample covariance of the true function with the perturbed target function $\tilde{y}$. $\tilde{\rho}_{c^*}$ is a useful reference against which to assess filter performance, in the sense that a filter that reproduces the same signal-to-noise ratio as the input data would achieve $\rho_{[1]} - \tilde{\rho}_{c^*} = 0$. 

Fig.~\ref{fig:fig_stats}a shows the percentiles of the distribution of $\rho_{[1]} - \tilde{\rho}_{c^*}$ versus noise level $\sigma_R$ (the purple $50\%$-curve for example corresponds to the median of this distribution). For low noise levels, the ranking performs very well: In fact, as can be seen from the initial low-noise regime, where $\langle \rho_{[1]} - \tilde{\rho}_{c^*} \rangle_{50\%} > 0$, we can obtain models with signal-to-noise ratios that are effectively larger than those of the datasets used to construct the feature network. Looking beyond the top-ranked feature, we observe how covariance and confidence decay towards lower ranks (larger $c/C$): The median of the true covariance $\rho_{[c]}$ with the unperturbed targets experiences a drastic drop as the noise level increases beyond $\sigma_R = 1$ (see Fig.~\ref{fig:fig_stats}b). For lower noise levels, the top-ranked features still tend to be useful, with $\rho_{[c]} \gtrsim 0.8$. Beyond $\sigma_R > 1$, however, the predictive power of the highest-covariance feature is marginal at best. This information breakdown is successfully detected by the filtering (see Fig.~\ref{fig:fig_stats}c), with the median of the confidence $\langle q(\varphi_{[c]} \rangle$ rapidly dropping to $0.5$ for the highest-ranked features, indicating that these functions are no longer better than the random control and that signal extraction was unsuccessful.

\subsection{Molecular reactivity models}

We tailor the physicochemical representations to each reaction system individually so that they adequately reflect the respective reaction class~\cite{beker_prediction_2019}. We broadly distinguish between ``chemisorption'' and ``physisorption'' descriptors (see Table~\ref{tab:descriptors} of the Methods section for a more detailed summary): Chemisorption descriptors capture molecular behaviour when undergoing chemical transformations, in particular the breaking and formation of chemical bonds; physisorption descriptors address weaker binding modes and interactions, associated with, e.g., electrostatic and dispersive interactions, hydrogen bonding or solvation. Chemisorption descriptors therefore include: (vertical) electron affinities and ionization energies, reorganization energies, and vibrational intensities. Physisorption descriptors include: electrostatic properties such as partial charges and polar surface area, partition coefficients, solubilities, and hydrogen-bonding parameters.

The Buchwald-Hartwig system is placed primarily in the ``chemisorption'' regime regarding both the aryl halide and additive -- as opposed to drug-drug synergy, which is ultimately a ``physisorption'' effect given the nature of ligand-protein interactions. The asymmetric hydrogenation cannot be classified as easily, but regarding the role of the bisphosphine ligand, weaker binding modes most likely outweigh strong chemical associations in determining the diastereoselectivity of the ligand-catalyst-substrate complex. However, given the symmetry properties of diastereomers, considering only global properties of the molecule (such as the total partition coefficient or polar surface area) {\it cannot} be enough to model product selectivity (in fact, tests show that the network filtering then concludes that no significant feature can be detected). Instead, we need to capture the asymmetry of the ligand by forming ``symmetric'' ($x^+_i = x_i + x_i'$) and ``antisymmetric'' ($x^- = | x_i - x_i' |$) combinations of physisorption parameters $x_i$ and $x_i'$  obtained for each of the two phosphine fragments (R and R') of the ligand.  For the synergy dataset, an analogous procedure is used to appropriately encode molecular combinations so that the descriptor is invariant to the ordering of the molecules of the pair: For each molecular feature $x_i$ and $x_i'$ describing the individual molecules $A$ and $A'$ of a pair $(A,A')$, we incorporate both their sum $x_i^+$ and absolute difference $x_i^-$. 

\begin{figure*}[t]%[tbhp]
\centering
\includegraphics[width=1.0\linewidth]{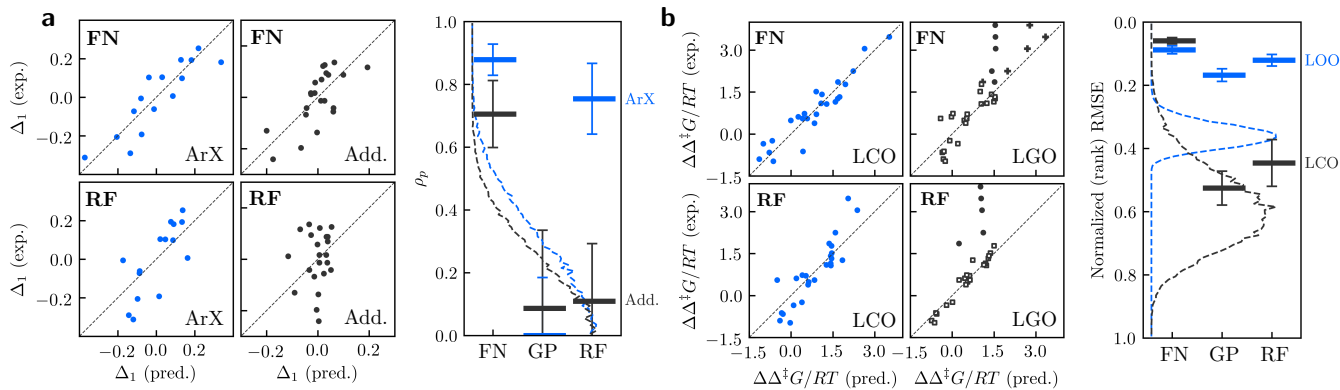}
\caption{Prediction performance for the (a) Buchwald-Hartwig and (b) asymmetric hydrogenation reaction. The scatter plots compare the predicted vs measured partial yield associated with the aryl halid and additive in the case of the Buchwald-Hartwig system, and the predicted vs measured ligand diastereoselectivity $\Delta \Delta^\ddagger G$ in the case of the asymmetric hydrogenation, with top panels showing the results for the feature network (FN), bottom panels for the random forest (RF). The benchmark in (b) considers two cross-validation approaches, leave-one-out (LOO) and leave-class-out (LCO) compared to only LOO validation for the Buchwald-Hartwig reaction. The LCO procedure uses the Walphos and Zhaophos scaffolds as test class, with test (training) predictions shown as filled (open) circles in the scatter plots. Finally, the bar plots compare the performance metrics for the Pearson correlation coefficient $\rho_p$ in panel (a), and (rank) RMSE for the LOO (LCO) test in panel (b), for models FN and RF, as well as the structural SOAP kernel (GP).}
\label{fig:fig_bench}
\end{figure*}

\subsubsection*{Network visualization} We will analyse the feature networks visually before benchmarking their performance. The networks that we construct here incorporate four layers (generations) of features: First, the root layer, $G_0$, consisting of the input descriptors; then, a layer of unary transformations, $G_1$ acting on the root layer; a layer of binary transformations, $G_2$ operating on the root layer; and, finally, a binary layer, $G_3$ acting on $G_0$ and $G_1$ combined. Visualizations of the networks are shown in Fig.~\ref{fig:fig_visual} for the (a) aryl-halide and (b) additive yield effect $\pyield_1$, (c) the ligand diastereoselectivity effect $\Delta \Delta^\ddagger G$, and (d) pair synergy label (i.e., classification of a molecular combination as synergistic vs non-synergistic). The correlation measure chosen is $\rho = \rho_p$ for (a)-(c) (regression tasks), and $\rho = \rho_\pm$ for network (d) (classification task). The organization of the graphs follows the schematic shown in the centre: Nodes are indicated with circles, links between nodes as arches. Root nodes are located on the innermost circle (as labelled by the dashed lines directed radially outwards), subsequent generations on concentric circles of increasing radii. The size and colouring of a node $\varphi_c$ and the links connecting this node to its parents follow from its normalized signal strength $|\hat{\rho}_c q_c| = |\rho_c q_c / \rho_\mathrm{max}|$, where $\rho_\mathrm{max}$ is the largest-magnitude correlation sampled by the network. Finally, the angle subtended by the root nodes is proportional to the aggregated signal strength of all its descendants in order to allocate more space to physically significant features.

The feature networks display interesting structuring in the form of ``pivots'' and ``resonances'': Pivots here refer to feature nodes that produce strong signals across a large number of descended features. For the aryl-halide, for example, the unary transformations of the reorganization energy $\lambda_h$ ($h$ for hole as opposed to electron) make up such a pivot. A second, particularly pronounced pivot is due to unary transformations of the antisymmetric $\log(P)^-$ in graph (c), with virtually all the signal stored in the network being descended from this descriptor. Resonances on the other hand are binary feature nodes descended from features that are relatively insignificant when considered individually, but give rise to a strong signal when combined: Examples here are the combinations of the partial charges $q_{H2}$ and $q_{H3}$ of the aryl halide, and $q_{C3}$ and $q_{O1}$ of the additive. This is consistent with the idea that charge pairs of this type can act as sensors which detect charge flow and charge reorganization and thus changes in reactivity caused by the addition of functional groups. It is noteworthy that the pivots and resonances descended from electronic observables such as reorganization energies and electron affinities on the one hand, and electrostatic observables such as partial charges on the other hand, coexist largely independently of each other, with very little cross-over between them. Among the top-ranked nodes of network (a), for example, we obtain the ``electronic'' feature $\varphi = \mathrm{IE}_v  \lambda_h^2$ with a signal strength of $\rho q = -0.87 \pm 0.05$ compared to $-0.91 \pm 0.06$ for the electrostatic feature $\varphi = \sqrt{q_{H2}} + \sqrt{q_{H3}}$.

The feature network describing the ligand effect on diastereoselectivity indicates a strong preference for antisymmetric features $x^-$ as opposed to symmetric features $x^+$ (see Fig.~\ref{fig:fig_visual}c). In fact, when leaving out all antisymmetric base features, the entire signal within the network fades. This can again be rationalized by the symmetries underlying diastereoselectivity, which should indeed only be captured by physicochemical descriptors that impose a sense of direction on the molecule. Due the $\log(P)^-$ pivot, the ranking of features of the network is relatively dense at the top end. It is in situations like this that the covariance attribution (see SI appendix) is useful: To first order, $\varphi_{[1]} = \sqrt{\log(P)^- Q_C^-}$ with a signal strength $\rho q = 0.86 \pm 0.05$ appears at the top ($Q_C$ is a charge asymmetry measure on the PC$_3$ group). The feature $\varphi_{[2]} = \log(P)^- + \mathrm{ln}(f_<^-)$ appears in second place, with $\rho q = 0.85 \pm 0.04$ ($f_<$ measures the free volume around the phosphorus centres). Inspection of the covariance contributions, however, reveals that $f_<^-$ contributes around $53\%$ to the covariance signal, compared to less than $20\%$ contributed by $Q_C$, therefore rendering $\varphi_{[2]}$ the more robust feature.

The synergy feature network (Fig.~\ref{fig:fig_visual}d), by contrast, is dominated by features derived from symmetric inputs -- in particular, the total number of hydrogen acceptors $n_a^+$ and added solubility $\sigma^+$ of the molecular pairs. The top-ranked feature is $\varphi = (n_a^+ \sigma^+)^2$ with a signal strength of $0.50 \pm 0.02$ and confidence $q > 0.999$. The fact that nodes descended from $x^+$ descriptors carry significantly more signal than those descended from $x^-$ descriptors indicates that synergy is more likely to occur if the molecules are physicochemically complementary and target related nodes within a cellular process. A point of concern here is, however, that the importance of solubility specifically could also indicate a measurement artifact: As the synergy score is evaluated based on concentration matrices, solubility effects could, for example, cause a discrepancy between actual and assumed concentration and this way confound the determined score.\\

\subsection{Predictive modelling} The diligent filtering and bootstrapping procedure that is part of the feature network analysis in principle allows assessing model viability without cross-validation. Below we show that, even when subjected to challenging cross-validation rules, the signal detected by the network analysis persists. During this cross-validation, the networks are constructed and evaluated using only a subset of the data, and least-squares ensemble regression on the top-ranked feature is used to make predictions for the test samples withheld during the training.

We will compare the quality of the predictions achieved by the feature-network approach (FN) to that of a Gaussian process (GP) and random forest (RF). The Gaussian processes trained here use the structural SOAP kernel~\cite{bartok_representing_2013}, which has been shown to achieve superior performance in inferring a large range of molecular properties~\cite{de_comparing_2016,bartok_machine_2017}. This kernel is designed to arrive at predictions via smooth substructure matching: Here we use it as a baseline that estimates how well a (by design) structurally-biased machine-learning framework performs. Random forests on the other hand are often the best-performing models in molecular informatics applications, due to their ability to overcome sample bias and model even highly nonlinear relationships, while being extremely parameter-efficient.

For the Buchwald-Hartwig reaction, we use leave-compound-out (LCO) cross-validation to assess the regression performance for the partial yields $\pyield_1$ of the aryl halide (ArX) and additive. Correlation plots pertaining to the feature network and random forest are shown in Fig.~\ref{fig:fig_bench}a, top and bottom panel, respectively. The correlation achieved by the FN model has a clear margin over the random forest, as quantified by the Pearson correlation coefficients indicated in the right-hand panel. Especially in the case of the additive effect, the performance differs drastically, with the RF not performing in any way better than the random control (see the distribution of null correlations indicated by the dashed curves). The vanishing predictive power of the GP model (the SOAP kernel) meanwhile confirms that the dataset is in both cases so sparse (but at the same time diverse) that structural interpolation fails entirely.

For the asymmetric hydrogenation, the models of ligand diastereoselectivity lead us to similar conclusions (Fig.~\ref{fig:fig_bench}b): Here we consider two different cross-validation schemes, a leave-compound-out and leave-group-out (LGO) procedure (see Fig.~\sifigcv~of the SI appendix for further cross-validation scenarios). The latter is a more challenging test case, in that we exclude the best-performing family of ligands (containing the Walphos scaffold, see Fig.~\ref{fig:fig_reactions}b) from the training, so that the test ligands all exhibit a higher diastereoselectivity than any of the training compounds. Moreover, we add one bisphosphine ligand to the test which has been reported to have $> 90\%$ diastereoselectivity~\cite{han_asymmetric_2017}, but was not part of our own experimental campaign. Whereas the models all perform adequately for LCO cross-validation (as quantified by the test RMSE), the performance in ranking the test samples with respect to the training set (here quantified by a relative rank RMSE) differs drastically, with only the FN model displaying a distinctly better-than-random performance. In fact, even though the absolute performance is relatively poor across all models, the feature network correctly assigns top ranks to the Walphos and Zhaophos ligands relative to the training data (see the filled black circles in Fig.~\ref{fig:fig_bench}b indicating the test predictions vs the open circles indicating training predictions). A significant improvement results if we include linear base features in the regression (black crosses), in which case the model extrapolates remarkably well to unseen compounds.

\begin{figure*}[t]%[tbhp]
\centering
\includegraphics[width=1.0\linewidth]{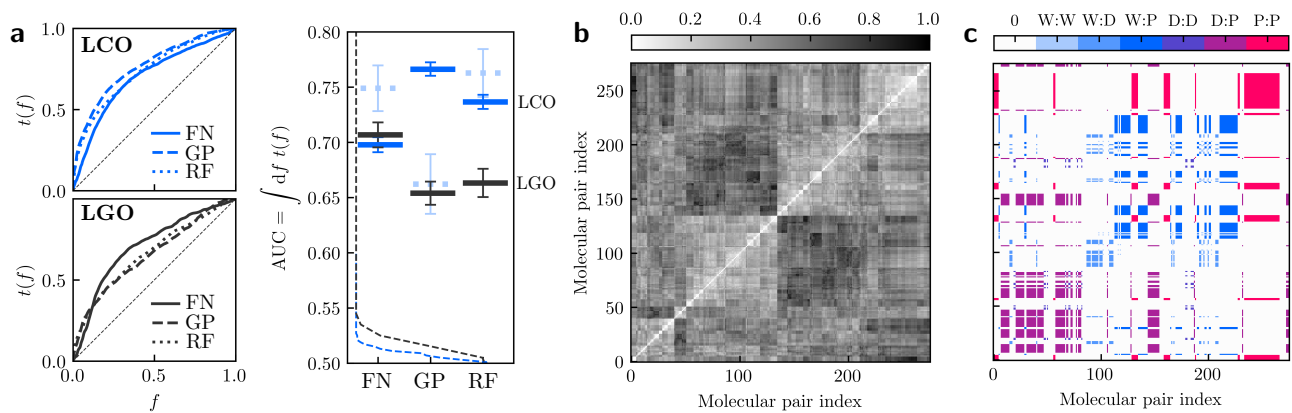}
\caption{(a) Predictions of drug synergy: The left-hand panel shows the receiver operating characteristics for two test categories resembling leave-compound-out (LCO) and leave-group-out (LGO) cross-validation. A third test category includes combinations measured at a second screening site, with the combinations again overlapping with the training set with exactly one compound (see dotted lines in AUC plot). The AUCs obtained by the models for the three different test situations are compared on the right-hand side, relative to the AUC distribution of a random null model (dashed lines). (b) Clustering view of the similarity matrix comparing antimicrobial combinations based on the feature network distance metric $d_{ab}$, resulting in (c) distinct structuring upon projection of the pair matrix onto antibiotic action modes (W: cell wall synthesis, D: dna replication/transcription, P: protein synthesis -- with, for example, a pair label P:P indicating that both partners target protein synthesis). A matrix entry is coloured in if both pairs linked to that entry have the same action mode X:Y. }
\label{fig:fig_synergy}
\end{figure*}

Finally, for the synergy classification, we consider ten splits of the data collected by one of the NCI-ALMANAC screening sites: For every split, we remove all molecular combinations containing one of 40 randomly selected test compounds from the training. The test combinations are therefore split onto two partitions: A partition resembling LCO cross-validation, where only one compound featuring in the combination was seen during the training; and a second partition resembling LGO cross-validation, where {\it neither} of the two compounds was seen during the training. The latter scenario is a significantly more rigorous test of a model's ability to generalize. Fig.~\ref{fig:fig_synergy}a shows the receiver operating charateristics for the LCO (blue set) and LGO partition (black set). The corresponding AUCs are compared in the right-hand panel: We find that only the feature network performs similarly for both test categories, whereas the Gaussian process and random forest appear to outperform the feature network regarding LCO, but then fall behind when subjected to the more challenging LGO set. This discrepancy between the performance on a ``close'' test set (LCO) and ``remote'' test set (LGO) is symptomatic of a structurally biased model. We note that a previous modelling study of the NCI-ALMANAC dataset based primarily on structural descriptors excludes LGO-type tests on genuinely new combinations~\cite{sidorov_predicting_2018}. The results here show, however, that these tests are crucial in assessing model bias and transferability.

\subsection{Mechanistic clustering} In this final section we propose an unsupervised route how feature networks can be leveraged for mechanistic insight. The supervised models we have considered so far are -- though sampled from a high-dimensional nonlinear space -- ultimately extremely reductive: The information captured by simple nonlinear correlations can be greater than intuition expects, but it is questionable whether the feature-network approach can succeed if the microscopic pathways driving a molecular reaction are diverse -- and if the rules that determine which molecular combination follows which of these pathways are difficult to infer from the available data and chosen descriptors. An unsupervised approach can then help us extract information about potentially coexisting reaction pathways and bottlenecks: The idea is to cluster the data samples into mechanistic categories based on their covariance behaviour sampled across a large number of derived features. To implement this idea, we decompose the covariance signal $\rho_c$ between a function $\varphi_c$ and $y$ onto contributions from the $N$ data samples: $\rho_c = \sum_{a=1}^N \Delta \rho_{ac}$ (see the SI appendix for details). We then construct a distance measure $d_{ab}$ capturing whether two samples $a$ and $b$ tend to come down on the same side ($\Delta \rho_{ac} \Delta \rho_{bc} > 0$) or on opposite sides ($\Delta \rho_{ac} \Delta \rho_{bc} < 0$) of the feature-target correlations,
\inserteq{
  d_{ab} = \frac{1}{C} \sum_{c=1}^C \frac{1}{2}\left( 1 - \frac{\Delta \rho_{ac} \Delta \rho_{bc}}{|\Delta \rho_{ac}| |\Delta \rho_{bc} |} \right),
}
where the sum extends over a suitable subset of the feature nodes. When tested on synthetic data (see Fig.~{\sifigsyn} of the SI appendix), where half the samples follow a randomly generated rule $y_a = \varphi(\bm{x}_a)$, half the samples a rule $y_a' = \varphi'(\bm{x}_a)$, hierarchical clustering on the distance metric $d_{ab}$ reveals that clusters tend to group together samples that are mechanistically related -- despite the descriptors $\bm{x}$ being sampled from exactly the same distribution and hence not providing any indication regarding their mechanistic class ($\varphi$ or $\varphi'$).

To carry this idea over to a real-life application, we use a published dataset on the antimicrobial synergy of 24 antibiotics~\cite{mason_prediction_2017}. A plot of the distance matrix $[D]_{A,B} = d_{AB}$ for the 276 unique antimicrobial combinations $A = (a,a')$ is shown in Fig.~\ref{fig:fig_synergy}b, where the order of the pairs is determined by the hierarchical clustering procedure. Given that many antibiotics fall into one of three mode-of-action categories -- targeting cell-wall synthesis (W), protein synthesis (P) or DNA replication/transcription (D) -- we can project the different mechanistic modes X:Y of a molecular pair $(a,a')$ (where, e.g., X = Y = P if both compounds $a$ and $a'$ target protein synthesis) onto this matrix plot, see Fig.~\ref{fig:fig_synergy}c. The matrix entries are colour-coded as indicated if the action modes of the two pairs pertaining to that entry match, and left white otherwise. This projection of experimental action modes on the hierarchical clustering matrix $D$ shows considerable non-random structuring, indicating an intriguing route how feature networks can be applied to gain insight into mechanistic sample similarity when left unsupervised.

\section{Conclusions}

The feature network analysis proposed in this work is a deep filtering technique designed to detect and verify even weak signals in sparse, noisy and nonlinear data. Applications to reaction systems including cross-coupling reactions, asymmetric catalysis and drug synergy indicate that the solutions proposed by the network reduce bias and improve extrapolation in challenging test cases. The models are easily interpretable: Visualization of the feature networks allows assessing the proposed solutions in context, control for base-variable correlations and derive physicochemical pharmacophores. Additionally, signal decomposition techniques identify redundant covariables in nonlinear features. In an unsupervised setting, the bulk of nonlinear features generated by the network can be used to quantify mechanistic similarity among the data samples.

For the asymmetric hydrogenation reaction studied here, the feature networks point to the importance of lipophilic gradients built into the molecular structure in driving diastereoselectivity. The reaction yield of the Buchwald-Hartwig amination is found to be affected strongly by subtle shifts in the electronic structure of the substrate and additive. The emphasis placed on symmetric features in the feature network models of anti-cancer activity indicates that synergistic interaction is more likely if the constituents of a molecular cocktail are physicochemically complementary.

In addition to making the modelling fully transparent, the framework displays excellent performance for large combinatorial datasets (even if they are ulteriorly sparse as shown here for the Buchwald-Hartwig amination). A particularly important application scenario could be in guiding dataset construction during the initial phases of high-noise experimental screens. Future work needs to reveal to which degree the wealth of statistical information embedded in the networks can also be exploited in modelling dense, low-noise datasets, where parameter-heavy approaches tend to dominate.

Next to predictive modelling, feature network filtering could also be used to control for undesired experimental artifacts by searching for signals derived from experimental control variables (e.g., solubilitiy, or the resolution and B-factors in a ligand binding screen). Integration with high-throughput experimentation would then have an advantage in both guiding exploration of chemical space and providing feedback on potential experimental confounds that warrant a follow-up investigation.

\subsection*{Supporting Information (SI) and Software} The SI Appendix includes details on the yield deconvolution, covariance attribution, descriptor calculation and predictive modelling. The feature network filter is implemented in the open-source software package \texttt{soap++}: The code and examples are available at github.com/capoe/soapxx. Experimental data will be made available at the University of Cambridge Data Repository, doi:10.17863/cam.XXXX.

%\matmethods{ 
\section*{Materials and Methods}

\subsection*{Experimental procedure} Experiments were conducted in an argon-filled Vigor glovebox in a stainless steel screening autoclave (Cat7 by HEL), using $\unit[10]{mL}$ glass reaction vials. The substrate, catalyst precursor and ligand were weighed and charged into the reactor, followed by the solvent and a magnetic stirrer. The autoclave was sealed and purged with H$_2$ three times, before being pressurised slowly to \unit[10]{barg} and heated to \unit[70]{$^\circ$C}. The stirring rate was 1000 rpm, and material loadings were \unit[0.1]{M}, 1\% Rh(CO)2(acac), 1.2\% ligand. Reaction time was \unit[17]{h} for all experiments. Conversion and diastereomeric excess (d.e.) were determined by chiral HPLC (Shimadzu Prominence, Chromspher column by Agilent, 8 min run time, 1 mL min-1 flow rate, \unit[22]{$^\circ$C} column temperature, acetonitrile : H2SO4 (98:2 v/v \%) mobile phase).
All models consider diastereoselectivity in terms of the normalised Gibbs free energy between the two product diastereomeric transition states, as calculated from transition state theory: $\Delta \Delta G^\ddagger = - RT \ln [(1-\textrm{d.e.})/(1+\textrm{d.e.})]$.
%\inserteq{
%\Delta \Delta G^\ddagger = - RT \ln \left(\frac{1-\textrm{d.e.}}{1+\textrm{d.e.}}\right).
%} % end eq

%In a previous operando NMR study we showed that this reaction undergoes the monohydride mechanism. [25] We have further investigated this mechanism by DFT to understand how the $\gamma$-lactam substrate is expected to bind to the catalyst-ligand species, and which intermediate along the reaction pathway is rate- and diastereo-determining. From the optimised intermediates, we found that the pathway through the carbo-Rh intermediates is preferred over that through the enolate intermediates (see  Supporting Information), and that the hydride insertion is rate- and enantio-determining, as shown in Figure 2. The electronically asymmetric nature of the substrate plays an important role here, with one side containing a polar moiety, and one non-polar side containing the propyl moiety. The orientation of the substrate during the first hydride insertion makes its interaction with the catalyst-substrate complex non-equivalent with respect to either of the two phosphine sections of the bidentate complex.

\subsection*{Quantum-chemical calculations} DFT calculations for the derivation of electronic-structure descriptors and vibrational IR intensities were carried out at the B3LYP/6-31+g(d,p) level of theory. Partial charges were fitted using the ChElPG approach as implemented in the Gaussian09 software suite. See the SI Appendix on more detailed information regarding the estimation of reorganization energies and the self-consistent best-match mode assignment used to detect shared vibrational modes.\\

\begin{table}%[tbhp]
\centering
\begin{tabular}{ll}
%\midrule
%Symbol & Description \\
\multicolumn{2}{l}{ \textbf{Electrostatic} } \\ \hline
%\midrule
$q_*$ & partial charges \\
$\mu$ & molecular dipole \\
$Q_C$ & carbon charge asymm. $\sqrt{\sum_{i<j}^3 (q_i - q_j)^2}$ on PC$_3$ \\
$S_\pm$ & total polar surface area \\
$n_a$ & number of hydrogen-bond acceptors groups \\
$n_d$ & number of hydrogen-bond donor groups \\
\\ \multicolumn{2}{l}{ \textbf{Vibrational} } \\ \hline
%\midrule
$I_*$ & IR vibrational intensities of shared modes \\
\\ \multicolumn{2}{l}{ \textbf{Electronic} } \\ \hline
%\midrule
$\mathrm{EA}_v$ ($\mathrm{EA}_v'$) &  vertical EA for charging (discharging) \\
$\mathrm{IE}_v$ ($\mathrm{IE}_v'$) & vertical IE for charging (discharging) \\
$\lambda_h$ ($\lambda_e$) & reorganization energies for holes (electrons) \\
$\eta$ & hardness \\
$\chi$ & electronegativity \\
\\ \multicolumn{2}{l}{ \textbf{Thermodynamic} } \\ \hline
%\midrule
$\log(P)$ & partition coefficient \\
$S_j$ & surface $\log(P)$  contributions, $1 \leq j \leq 12$ \\
$\sigma$ & solubility \\
\\  \multicolumn{2}{l}{ \textbf{Steric} } \\ \hline
%\midrule
$f_<$ & free volume around phosphorus, $r_c = \unit[2.0]{\AA}$ \\
$f_>$ & free volume around phosphorus, $r_c = \unit[3.5]{\AA}$ \\
$M$ & molecular weight \\
%\bottomrule
\end{tabular}
\label{tab:descriptors}
\caption{Categories of physicochemical descriptors used in this work}
%\vspace{0.5cm}
\end{table}

\subsection*{Data processing} The deconvolution technique used to quantify the aryl-halide and additive effect on the Buchwald-Hartwig amination is described in the SI Appendix. 

The deconvolution of the solvent effect on the asymmetric hydrogenation showed that solvent performance is mostly determined by the solvent's dielectric and lipophilic properties. High-dielectric solvents were particularly detrimental to the reaction outcome. The ligand effect on diastereoselectivity was therefore measured with respect to 2-propanol as solvent, which could be shown to have near-neutral effect on both diastereoselectivity and conversion. As the measurements of diastereoselectivity at low yield are unreliable, ligands with a conversion outcome lower than 3\% were excluded from all models.

Synergy labels for the molecular combinations tested by the NCI-ALMANAC library were derived from the reported synergy scores $y$ (``ComboScore'') averaged over the individual cell lines. Bootstrapping is used to estimate the standard error $\sigma$ of this average $\langle y \rangle$. With $y > 0$ designating greater-than-additive efficacy, a molecular pair is labeled as synergistic only if $\langle y \rangle \geq 2 \sigma$ in order to exclude additive pairs from the class of synergistic combinations. Due to the small number of data samples ($< 4\%$) contributed by the screening centre at the Frederick National Laboratory for Cancer Research, as well as the different protocol used therein, only the data from the primary screening centres at the University of Pittsburgh and SRI International were considered.\\

\subsection*{Predictive modelling} The linear least-squares ensembles (LSEs) used by the FN regression were generated by resampling the raw residuals $\bm{\hat{\epsilon}} = \bm{y} - \bm{\hat{y}}$ obtained from a least-squares fit of the training data, where $\bm{y}$ and $\bm{\hat{y}}$ are the vectors of measured and predicted responses/targets, respectively. Each least-squares instance is hence trained with perturbed targets $\bm{\tilde{y}} = \bm{y} + \bm{\tilde{\epsilon}}$, where $\bm{\tilde{\epsilon}}$ is a vector with components bootstrapped from $\bm{\epsilon}$. The prediction outcome of the LSE is chosen as the median value to reduce sensitivity to outliers. An LSE size of around 2000 instances proved sufficient to achieve convergence. 

For the random-forest regressors (RFs), we considered the number of estimators $N$, tree depth $t$ and number of features for each split $f$ as hyperparameters. With $N \simeq 1000$ chosen large enough, the sensitivity to $t$ and $f$ proved small for the datasets given sufficiently deep trees with $t \geq 6$.

See the SI appendix for details on the hierarchical SOAP kernel used by the Gaussian process model.

\section*{Acknowledgements}
CP acknowledges funding from the European Union's Horizon 2020 research and innovation programme, Grant Agreement No. 676580 through the Novel Materials Discovery (NOMAD) Laboratory, a European Center of Excellence (https://
www.nomad-coe.eu). YA is grateful to UCB Pharma for funding his PhD studies. This research was in part supported by the National Research Foundation, Prime Minister's Office, Singapore under its CREATE programme. 

% Bibliography
\bibliography{literature,zotero,zotero_add,notes}

%merlin.mbs apsrev4-1.bst 2010-07-25 4.21a (PWD, AO, DPC) hacked
%Control: key (0)
%Control: author (8) initials jnrlst
%Control: editor formatted (1) identically to author
%Control: production of article title (-1) disabled
%Control: page (0) single
%Control: year (1) truncated
%Control: production of eprint (0) enabled
\begin{thebibliography}{40}%
\makeatletter
\providecommand \@ifxundefined [1]{%
 \@ifx{#1\undefined}
}%
\providecommand \@ifnum [1]{%
 \ifnum #1\expandafter \@firstoftwo
 \else \expandafter \@secondoftwo
 \fi
}%
\providecommand \@ifx [1]{%
 \ifx #1\expandafter \@firstoftwo
 \else \expandafter \@secondoftwo
 \fi
}%
\providecommand \natexlab [1]{#1}%
\providecommand \enquote  [1]{``#1''}%
\providecommand \bibnamefont  [1]{#1}%
\providecommand \bibfnamefont [1]{#1}%
\providecommand \citenamefont [1]{#1}%
\providecommand \href@noop [0]{\@secondoftwo}%
\providecommand \href [0]{\begingroup \@sanitize@url \@href}%
\providecommand \@href[1]{\@@startlink{#1}\@@href}%
\providecommand \@@href[1]{\endgroup#1\@@endlink}%
\providecommand \@sanitize@url [0]{\catcode `\\12\catcode `\$12\catcode
  `\&12\catcode `\#12\catcode `\^12\catcode `\_12\catcode `\%12\relax}%
\providecommand \@@startlink[1]{}%
\providecommand \@@endlink[0]{}%
\providecommand \url  [0]{\begingroup\@sanitize@url \@url }%
\providecommand \@url [1]{\endgroup\@href {#1}{\urlprefix }}%
\providecommand \urlprefix  [0]{URL }%
\providecommand \Eprint [0]{\href }%
\providecommand \doibase [0]{http://dx.doi.org/}%
\providecommand \selectlanguage [0]{\@gobble}%
\providecommand \bibinfo  [0]{\@secondoftwo}%
\providecommand \bibfield  [0]{\@secondoftwo}%
\providecommand \translation [1]{[#1]}%
\providecommand \BibitemOpen [0]{}%
\providecommand \bibitemStop [0]{}%
\providecommand \bibitemNoStop [0]{.\EOS\space}%
\providecommand \EOS [0]{\spacefactor3000\relax}%
\providecommand \BibitemShut  [1]{\csname bibitem#1\endcsname}%
\let\auto@bib@innerbib\@empty
%</preamble>
\bibitem [{\citenamefont {Isbrandt}\ \emph {et~al.}(2019)\citenamefont
  {Isbrandt}, \citenamefont {Sullivan},\ and\ \citenamefont
  {Newman}}]{isbrandt_high_2019}%
  \BibitemOpen
  \bibfield  {author} {\bibinfo {author} {\bibfnamefont {E.~S.}\ \bibnamefont
  {Isbrandt}}, \bibinfo {author} {\bibfnamefont {R.~J.}\ \bibnamefont
  {Sullivan}}, \ and\ \bibinfo {author} {\bibfnamefont {S.~G.}\ \bibnamefont
  {Newman}},\ }\href {\doibase 10.1002/anie.201812534} {\bibfield  {journal}
  {\bibinfo  {journal} {Angewandte Chemie International Edition}\ }\textbf
  {\bibinfo {volume} {58}},\ \bibinfo {pages} {7180} (\bibinfo {year}
  {2019})}\BibitemShut {NoStop}%
\bibitem [{\citenamefont {Granda}\ \emph {et~al.}(2018)\citenamefont {Granda},
  \citenamefont {Donina}, \citenamefont {Dragone}, \citenamefont {Long},\ and\
  \citenamefont {Cronin}}]{granda_controlling_2018}%
  \BibitemOpen
  \bibfield  {author} {\bibinfo {author} {\bibfnamefont {J.~M.}\ \bibnamefont
  {Granda}}, \bibinfo {author} {\bibfnamefont {L.}~\bibnamefont {Donina}},
  \bibinfo {author} {\bibfnamefont {V.}~\bibnamefont {Dragone}}, \bibinfo
  {author} {\bibfnamefont {D.-L.}\ \bibnamefont {Long}}, \ and\ \bibinfo
  {author} {\bibfnamefont {L.}~\bibnamefont {Cronin}},\ }\href {\doibase
  10.1038/s41586-018-0307-8} {\bibfield  {journal} {\bibinfo  {journal}
  {Nature}\ }\textbf {\bibinfo {volume} {559}},\ \bibinfo {pages} {377}
  (\bibinfo {year} {2018})}\BibitemShut {NoStop}%
\bibitem [{\citenamefont {Gromski}\ \emph {et~al.}(2019)\citenamefont
  {Gromski}, \citenamefont {Henson}, \citenamefont {Granda},\ and\
  \citenamefont {Cronin}}]{gromski_how_2019}%
  \BibitemOpen
  \bibfield  {author} {\bibinfo {author} {\bibfnamefont {P.~S.}\ \bibnamefont
  {Gromski}}, \bibinfo {author} {\bibfnamefont {A.~B.}\ \bibnamefont {Henson}},
  \bibinfo {author} {\bibfnamefont {J.~M.}\ \bibnamefont {Granda}}, \ and\
  \bibinfo {author} {\bibfnamefont {L.}~\bibnamefont {Cronin}},\ }\href
  {\doibase 10.1038/s41570-018-0066-y} {\bibfield  {journal} {\bibinfo
  {journal} {Nature Reviews Chemistry}\ }\textbf {\bibinfo {volume} {3}},\
  \bibinfo {pages} {119} (\bibinfo {year} {2019})}\BibitemShut {NoStop}%
\bibitem [{\citenamefont {Allen}\ \emph {et~al.}(2019)\citenamefont {Allen},
  \citenamefont {Leitch}, \citenamefont {Anson},\ and\ \citenamefont
  {Zajac}}]{allen_power_2019}%
  \BibitemOpen
  \bibfield  {author} {\bibinfo {author} {\bibfnamefont {C.~L.}\ \bibnamefont
  {Allen}}, \bibinfo {author} {\bibfnamefont {D.~C.}\ \bibnamefont {Leitch}},
  \bibinfo {author} {\bibfnamefont {M.~S.}\ \bibnamefont {Anson}}, \ and\
  \bibinfo {author} {\bibfnamefont {M.~A.}\ \bibnamefont {Zajac}},\ }\href
  {\doibase 10.1038/s41929-018-0220-4} {\bibfield  {journal} {\bibinfo
  {journal} {Nature Catalysis}\ }\textbf {\bibinfo {volume} {2}},\ \bibinfo
  {pages} {2} (\bibinfo {year} {2019})}\BibitemShut {NoStop}%
\bibitem [{\citenamefont {Zahrt}\ \emph {et~al.}(2019)\citenamefont {Zahrt},
  \citenamefont {Henle}, \citenamefont {Rose}, \citenamefont {Wang},
  \citenamefont {Darrow},\ and\ \citenamefont
  {Denmark}}]{zahrt_prediction_2019}%
  \BibitemOpen
  \bibfield  {author} {\bibinfo {author} {\bibfnamefont {A.~F.}\ \bibnamefont
  {Zahrt}}, \bibinfo {author} {\bibfnamefont {J.~J.}\ \bibnamefont {Henle}},
  \bibinfo {author} {\bibfnamefont {B.~T.}\ \bibnamefont {Rose}}, \bibinfo
  {author} {\bibfnamefont {Y.}~\bibnamefont {Wang}}, \bibinfo {author}
  {\bibfnamefont {W.~T.}\ \bibnamefont {Darrow}}, \ and\ \bibinfo {author}
  {\bibfnamefont {S.~E.}\ \bibnamefont {Denmark}},\ }\href {\doibase
  10.1126/science.aau5631} {\bibfield  {journal} {\bibinfo  {journal}
  {Science}\ }\textbf {\bibinfo {volume} {363}},\ \bibinfo {pages} {eaau5631}
  (\bibinfo {year} {2019})}\BibitemShut {NoStop}%
\bibitem [{\citenamefont {Reid}\ and\ \citenamefont
  {Sigman}(2018)}]{reid_comparing_2018}%
  \BibitemOpen
  \bibfield  {author} {\bibinfo {author} {\bibfnamefont {J.~P.}\ \bibnamefont
  {Reid}}\ and\ \bibinfo {author} {\bibfnamefont {M.~S.}\ \bibnamefont
  {Sigman}},\ }\href {\doibase 10.1038/s41570-018-0040-8} {\bibfield  {journal}
  {\bibinfo  {journal} {Nature Reviews Chemistry}\ }\textbf {\bibinfo {volume}
  {2}},\ \bibinfo {pages} {290} (\bibinfo {year} {2018})}\BibitemShut {NoStop}%
\bibitem [{\citenamefont {Rosales}\ \emph {et~al.}(2019)\citenamefont
  {Rosales}, \citenamefont {Wahlers}, \citenamefont {Lim{\'e}}, \citenamefont
  {Meadows}, \citenamefont {Leslie}, \citenamefont {Savin}, \citenamefont
  {Bell}, \citenamefont {Hansen}, \citenamefont {Helquist}, \citenamefont
  {Munday}, \citenamefont {Wiest},\ and\ \citenamefont
  {Norrby}}]{rosales_rapid_2019}%
  \BibitemOpen
  \bibfield  {author} {\bibinfo {author} {\bibfnamefont {A.~R.}\ \bibnamefont
  {Rosales}}, \bibinfo {author} {\bibfnamefont {J.}~\bibnamefont {Wahlers}},
  \bibinfo {author} {\bibfnamefont {E.}~\bibnamefont {Lim{\'e}}}, \bibinfo
  {author} {\bibfnamefont {R.~E.}\ \bibnamefont {Meadows}}, \bibinfo {author}
  {\bibfnamefont {K.~W.}\ \bibnamefont {Leslie}}, \bibinfo {author}
  {\bibfnamefont {R.}~\bibnamefont {Savin}}, \bibinfo {author} {\bibfnamefont
  {F.}~\bibnamefont {Bell}}, \bibinfo {author} {\bibfnamefont {E.}~\bibnamefont
  {Hansen}}, \bibinfo {author} {\bibfnamefont {P.}~\bibnamefont {Helquist}},
  \bibinfo {author} {\bibfnamefont {R.~H.}\ \bibnamefont {Munday}}, \bibinfo
  {author} {\bibfnamefont {O.}~\bibnamefont {Wiest}}, \ and\ \bibinfo {author}
  {\bibfnamefont {P.-O.}\ \bibnamefont {Norrby}},\ }\href {\doibase
  10.1038/s41929-018-0193-3} {\bibfield  {journal} {\bibinfo  {journal} {Nature
  Catalysis}\ }\textbf {\bibinfo {volume} {2}},\ \bibinfo {pages} {41}
  (\bibinfo {year} {2019})}\BibitemShut {NoStop}%
\bibitem [{\citenamefont {Segler}\ \emph {et~al.}(2018)\citenamefont {Segler},
  \citenamefont {Preuss},\ and\ \citenamefont {Waller}}]{segler_planning_2018}%
  \BibitemOpen
  \bibfield  {author} {\bibinfo {author} {\bibfnamefont {M.~H.~S.}\
  \bibnamefont {Segler}}, \bibinfo {author} {\bibfnamefont {M.}~\bibnamefont
  {Preuss}}, \ and\ \bibinfo {author} {\bibfnamefont {M.~P.}\ \bibnamefont
  {Waller}},\ }\href {\doibase 10.1038/nature25978} {\bibfield  {journal}
  {\bibinfo  {journal} {Nature}\ }\textbf {\bibinfo {volume} {555}},\ \bibinfo
  {pages} {604} (\bibinfo {year} {2018})}\BibitemShut {NoStop}%
\bibitem [{\citenamefont {Zhou}\ \emph {et~al.}(2017)\citenamefont {Zhou},
  \citenamefont {Li},\ and\ \citenamefont {Zare}}]{zhou_optimizing_2017}%
  \BibitemOpen
  \bibfield  {author} {\bibinfo {author} {\bibfnamefont {Z.}~\bibnamefont
  {Zhou}}, \bibinfo {author} {\bibfnamefont {X.}~\bibnamefont {Li}}, \ and\
  \bibinfo {author} {\bibfnamefont {R.~N.}\ \bibnamefont {Zare}},\ }\href
  {\doibase 10.1021/acscentsci.7b00492} {\bibfield  {journal} {\bibinfo
  {journal} {ACS Central Science}\ }\textbf {\bibinfo {volume} {3}},\ \bibinfo
  {pages} {1337} (\bibinfo {year} {2017})}\BibitemShut {NoStop}%
\bibitem [{\citenamefont {Chen}\ \emph {et~al.}(2018)\citenamefont {Chen},
  \citenamefont {Engkvist}, \citenamefont {Wang}, \citenamefont {Olivecrona},\
  and\ \citenamefont {Blaschke}}]{chen_rise_2018}%
  \BibitemOpen
  \bibfield  {author} {\bibinfo {author} {\bibfnamefont {H.}~\bibnamefont
  {Chen}}, \bibinfo {author} {\bibfnamefont {O.}~\bibnamefont {Engkvist}},
  \bibinfo {author} {\bibfnamefont {Y.}~\bibnamefont {Wang}}, \bibinfo {author}
  {\bibfnamefont {M.}~\bibnamefont {Olivecrona}}, \ and\ \bibinfo {author}
  {\bibfnamefont {T.}~\bibnamefont {Blaschke}},\ }\href {\doibase
  10.1016/j.drudis.2018.01.039} {\bibfield  {journal} {\bibinfo  {journal}
  {Drug Discovery Today}\ }\textbf {\bibinfo {volume} {23}},\ \bibinfo {pages}
  {1241} (\bibinfo {year} {2018})}\BibitemShut {NoStop}%
\bibitem [{\citenamefont {Ekins}\ \emph {et~al.}(2019)\citenamefont {Ekins},
  \citenamefont {Puhl}, \citenamefont {Zorn}, \citenamefont {Lane},
  \citenamefont {Russo}, \citenamefont {Klein}, \citenamefont {Hickey},\ and\
  \citenamefont {Clark}}]{ekins_exploiting_2019}%
  \BibitemOpen
  \bibfield  {author} {\bibinfo {author} {\bibfnamefont {S.}~\bibnamefont
  {Ekins}}, \bibinfo {author} {\bibfnamefont {A.~C.}\ \bibnamefont {Puhl}},
  \bibinfo {author} {\bibfnamefont {K.~M.}\ \bibnamefont {Zorn}}, \bibinfo
  {author} {\bibfnamefont {T.~R.}\ \bibnamefont {Lane}}, \bibinfo {author}
  {\bibfnamefont {D.~P.}\ \bibnamefont {Russo}}, \bibinfo {author}
  {\bibfnamefont {J.~J.}\ \bibnamefont {Klein}}, \bibinfo {author}
  {\bibfnamefont {A.~J.}\ \bibnamefont {Hickey}}, \ and\ \bibinfo {author}
  {\bibfnamefont {A.~M.}\ \bibnamefont {Clark}},\ }\href {\doibase
  10.1038/s41563-019-0338-z} {\bibfield  {journal} {\bibinfo  {journal} {Nature
  Materials}\ }\textbf {\bibinfo {volume} {18}},\ \bibinfo {pages} {435}
  (\bibinfo {year} {2019})}\BibitemShut {NoStop}%
\bibitem [{\citenamefont {Bart{\'o}k}\ \emph {et~al.}(2017)\citenamefont
  {Bart{\'o}k}, \citenamefont {De}, \citenamefont {Poelking}, \citenamefont
  {Bernstein}, \citenamefont {Kermode}, \citenamefont {Cs{\'a}nyi},\ and\
  \citenamefont {Ceriotti}}]{bartok_machine_2017}%
  \BibitemOpen
  \bibfield  {author} {\bibinfo {author} {\bibfnamefont {A.~P.}\ \bibnamefont
  {Bart{\'o}k}}, \bibinfo {author} {\bibfnamefont {S.}~\bibnamefont {De}},
  \bibinfo {author} {\bibfnamefont {C.}~\bibnamefont {Poelking}}, \bibinfo
  {author} {\bibfnamefont {N.}~\bibnamefont {Bernstein}}, \bibinfo {author}
  {\bibfnamefont {J.~R.}\ \bibnamefont {Kermode}}, \bibinfo {author}
  {\bibfnamefont {G.}~\bibnamefont {Cs{\'a}nyi}}, \ and\ \bibinfo {author}
  {\bibfnamefont {M.}~\bibnamefont {Ceriotti}},\ }\href {\doibase
  10.1126/sciadv.1701816} {\bibfield  {journal} {\bibinfo  {journal} {Science
  Advances}\ }\textbf {\bibinfo {volume} {3}},\ \bibinfo {pages} {e1701816}
  (\bibinfo {year} {2017})}\BibitemShut {NoStop}%
\bibitem [{\citenamefont {Goldsmith}\ \emph {et~al.}(2018)\citenamefont
  {Goldsmith}, \citenamefont {Esterhuizen}, \citenamefont {Liu}, \citenamefont
  {Bartel},\ and\ \citenamefont {Sutton}}]{goldsmith_machine_2018}%
  \BibitemOpen
  \bibfield  {author} {\bibinfo {author} {\bibfnamefont {B.~R.}\ \bibnamefont
  {Goldsmith}}, \bibinfo {author} {\bibfnamefont {J.}~\bibnamefont
  {Esterhuizen}}, \bibinfo {author} {\bibfnamefont {J.-X.}\ \bibnamefont
  {Liu}}, \bibinfo {author} {\bibfnamefont {C.~J.}\ \bibnamefont {Bartel}}, \
  and\ \bibinfo {author} {\bibfnamefont {C.}~\bibnamefont {Sutton}},\ }\href
  {\doibase 10.1002/aic.16198} {\bibfield  {journal} {\bibinfo  {journal}
  {AIChE Journal}\ }\textbf {\bibinfo {volume} {64}},\ \bibinfo {pages} {2311}
  (\bibinfo {year} {2018})}\BibitemShut {NoStop}%
\bibitem [{\citenamefont {Meyer}\ \emph {et~al.}(2018)\citenamefont {Meyer},
  \citenamefont {Sawatlon}, \citenamefont {Heinen}, \citenamefont {von
  Lilienfeld},\ and\ \citenamefont {Corminboeuf}}]{meyer_machine_2018}%
  \BibitemOpen
  \bibfield  {author} {\bibinfo {author} {\bibfnamefont {B.}~\bibnamefont
  {Meyer}}, \bibinfo {author} {\bibfnamefont {B.}~\bibnamefont {Sawatlon}},
  \bibinfo {author} {\bibfnamefont {S.}~\bibnamefont {Heinen}}, \bibinfo
  {author} {\bibfnamefont {O.~A.}\ \bibnamefont {von Lilienfeld}}, \ and\
  \bibinfo {author} {\bibfnamefont {C.}~\bibnamefont {Corminboeuf}},\ }\href
  {\doibase 10.1039/C8SC01949E} {\bibfield  {journal} {\bibinfo  {journal}
  {Chemical Science}\ }\textbf {\bibinfo {volume} {9}},\ \bibinfo {pages}
  {7069} (\bibinfo {year} {2018})}\BibitemShut {NoStop}%
\bibitem [{\citenamefont {Raccuglia}\ \emph {et~al.}(2016)\citenamefont
  {Raccuglia}, \citenamefont {Elbert}, \citenamefont {Adler}, \citenamefont
  {Falk}, \citenamefont {Wenny}, \citenamefont {Mollo}, \citenamefont {Zeller},
  \citenamefont {Friedler}, \citenamefont {Schrier},\ and\ \citenamefont
  {Norquist}}]{raccuglia_machine-learning-assisted_2016}%
  \BibitemOpen
  \bibfield  {author} {\bibinfo {author} {\bibfnamefont {P.}~\bibnamefont
  {Raccuglia}}, \bibinfo {author} {\bibfnamefont {K.~C.}\ \bibnamefont
  {Elbert}}, \bibinfo {author} {\bibfnamefont {P.~D.~F.}\ \bibnamefont
  {Adler}}, \bibinfo {author} {\bibfnamefont {C.}~\bibnamefont {Falk}},
  \bibinfo {author} {\bibfnamefont {M.~B.}\ \bibnamefont {Wenny}}, \bibinfo
  {author} {\bibfnamefont {A.}~\bibnamefont {Mollo}}, \bibinfo {author}
  {\bibfnamefont {M.}~\bibnamefont {Zeller}}, \bibinfo {author} {\bibfnamefont
  {S.~A.}\ \bibnamefont {Friedler}}, \bibinfo {author} {\bibfnamefont
  {J.}~\bibnamefont {Schrier}}, \ and\ \bibinfo {author} {\bibfnamefont
  {A.~J.}\ \bibnamefont {Norquist}},\ }\href {\doibase 10.1038/nature17439}
  {\bibfield  {journal} {\bibinfo  {journal} {Nature}\ }\textbf {\bibinfo
  {volume} {533}},\ \bibinfo {pages} {73} (\bibinfo {year} {2016})}\BibitemShut
  {NoStop}%
\bibitem [{\citenamefont {Coley}\ \emph {et~al.}(2019)\citenamefont {Coley},
  \citenamefont {Jin}, \citenamefont {Rogers}, \citenamefont {Jamison},
  \citenamefont {Jaakkola}, \citenamefont {Green}, \citenamefont {Barzilay},\
  and\ \citenamefont {Jensen}}]{coley_graph-convolutional_2019}%
  \BibitemOpen
  \bibfield  {author} {\bibinfo {author} {\bibfnamefont {C.~W.}\ \bibnamefont
  {Coley}}, \bibinfo {author} {\bibfnamefont {W.}~\bibnamefont {Jin}}, \bibinfo
  {author} {\bibfnamefont {L.}~\bibnamefont {Rogers}}, \bibinfo {author}
  {\bibfnamefont {T.~F.}\ \bibnamefont {Jamison}}, \bibinfo {author}
  {\bibfnamefont {T.~S.}\ \bibnamefont {Jaakkola}}, \bibinfo {author}
  {\bibfnamefont {W.~H.}\ \bibnamefont {Green}}, \bibinfo {author}
  {\bibfnamefont {R.}~\bibnamefont {Barzilay}}, \ and\ \bibinfo {author}
  {\bibfnamefont {K.~F.}\ \bibnamefont {Jensen}},\ }\href {\doibase
  10.1039/C8SC04228D} {\bibfield  {journal} {\bibinfo  {journal} {Chemical
  Science}\ }\textbf {\bibinfo {volume} {10}},\ \bibinfo {pages} {370}
  (\bibinfo {year} {2019})}\BibitemShut {NoStop}%
\bibitem [{\citenamefont {Ahneman}\ \emph {et~al.}(2018)\citenamefont
  {Ahneman}, \citenamefont {Estrada}, \citenamefont {Lin}, \citenamefont
  {Dreher},\ and\ \citenamefont {Doyle}}]{ahneman_predicting_2018}%
  \BibitemOpen
  \bibfield  {author} {\bibinfo {author} {\bibfnamefont {D.~T.}\ \bibnamefont
  {Ahneman}}, \bibinfo {author} {\bibfnamefont {J.~G.}\ \bibnamefont
  {Estrada}}, \bibinfo {author} {\bibfnamefont {S.}~\bibnamefont {Lin}},
  \bibinfo {author} {\bibfnamefont {S.~D.}\ \bibnamefont {Dreher}}, \ and\
  \bibinfo {author} {\bibfnamefont {A.~G.}\ \bibnamefont {Doyle}},\ }\href
  {\doibase 10.1126/science.aar5169} {\bibfield  {journal} {\bibinfo  {journal}
  {Science}\ }\textbf {\bibinfo {volume} {360}},\ \bibinfo {pages} {186}
  (\bibinfo {year} {2018})}\BibitemShut {NoStop}%
\bibitem [{\citenamefont {Sigman}\ \emph {et~al.}(2016)\citenamefont {Sigman},
  \citenamefont {Harper}, \citenamefont {Bess},\ and\ \citenamefont
  {Milo}}]{sigman_development_2016}%
  \BibitemOpen
  \bibfield  {author} {\bibinfo {author} {\bibfnamefont {M.~S.}\ \bibnamefont
  {Sigman}}, \bibinfo {author} {\bibfnamefont {K.~C.}\ \bibnamefont {Harper}},
  \bibinfo {author} {\bibfnamefont {E.~N.}\ \bibnamefont {Bess}}, \ and\
  \bibinfo {author} {\bibfnamefont {A.}~\bibnamefont {Milo}},\ }\href {\doibase
  10.1021/acs.accounts.6b00194} {\bibfield  {journal} {\bibinfo  {journal}
  {Accounts of Chemical Research}\ }\textbf {\bibinfo {volume} {49}},\ \bibinfo
  {pages} {1292} (\bibinfo {year} {2016})}\BibitemShut {NoStop}%
\bibitem [{\citenamefont {G{\'o}mez-Bombarelli}\ \emph
  {et~al.}(2018)\citenamefont {G{\'o}mez-Bombarelli}, \citenamefont {Wei},
  \citenamefont {Duvenaud}, \citenamefont {Hern{\'a}ndez-Lobato}, \citenamefont
  {S{\'a}nchez-Lengeling}, \citenamefont {Sheberla}, \citenamefont
  {Aguilera-Iparraguirre}, \citenamefont {Hirzel}, \citenamefont {Adams},\ and\
  \citenamefont {Aspuru-Guzik}}]{gomez-bombarelli_automatic_2018}%
  \BibitemOpen
  \bibfield  {author} {\bibinfo {author} {\bibfnamefont {R.}~\bibnamefont
  {G{\'o}mez-Bombarelli}}, \bibinfo {author} {\bibfnamefont {J.~N.}\
  \bibnamefont {Wei}}, \bibinfo {author} {\bibfnamefont {D.}~\bibnamefont
  {Duvenaud}}, \bibinfo {author} {\bibfnamefont {J.~M.}\ \bibnamefont
  {Hern{\'a}ndez-Lobato}}, \bibinfo {author} {\bibfnamefont {B.}~\bibnamefont
  {S{\'a}nchez-Lengeling}}, \bibinfo {author} {\bibfnamefont {D.}~\bibnamefont
  {Sheberla}}, \bibinfo {author} {\bibfnamefont {J.}~\bibnamefont
  {Aguilera-Iparraguirre}}, \bibinfo {author} {\bibfnamefont {T.~D.}\
  \bibnamefont {Hirzel}}, \bibinfo {author} {\bibfnamefont {R.~P.}\
  \bibnamefont {Adams}}, \ and\ \bibinfo {author} {\bibfnamefont
  {A.}~\bibnamefont {Aspuru-Guzik}},\ }\href {\doibase
  10.1021/acscentsci.7b00572} {\bibfield  {journal} {\bibinfo  {journal} {ACS
  Central Science}\ }\textbf {\bibinfo {volume} {4}},\ \bibinfo {pages} {268}
  (\bibinfo {year} {2018})}\BibitemShut {NoStop}%
\bibitem [{\citenamefont {Harper}\ and\ \citenamefont
  {Sigman}(2011)}]{harper_predicting_2011}%
  \BibitemOpen
  \bibfield  {author} {\bibinfo {author} {\bibfnamefont {K.~C.}\ \bibnamefont
  {Harper}}\ and\ \bibinfo {author} {\bibfnamefont {M.~S.}\ \bibnamefont
  {Sigman}},\ }\href {\doibase 10.1073/pnas.1013331108} {\bibfield  {journal}
  {\bibinfo  {journal} {Proceedings of the National Academy of Sciences}\
  }\textbf {\bibinfo {volume} {108}},\ \bibinfo {pages} {2179} (\bibinfo {year}
  {2011})}\BibitemShut {NoStop}%
\bibitem [{\citenamefont {Milo}\ \emph {et~al.}(2015)\citenamefont {Milo},
  \citenamefont {Neel}, \citenamefont {Toste},\ and\ \citenamefont
  {Sigman}}]{milo_data-intensive_2015}%
  \BibitemOpen
  \bibfield  {author} {\bibinfo {author} {\bibfnamefont {A.}~\bibnamefont
  {Milo}}, \bibinfo {author} {\bibfnamefont {A.~J.}\ \bibnamefont {Neel}},
  \bibinfo {author} {\bibfnamefont {F.~D.}\ \bibnamefont {Toste}}, \ and\
  \bibinfo {author} {\bibfnamefont {M.~S.}\ \bibnamefont {Sigman}},\ }\href
  {\doibase 10.1126/science.1261043} {\bibfield  {journal} {\bibinfo  {journal}
  {Science}\ }\textbf {\bibinfo {volume} {347}},\ \bibinfo {pages} {737}
  (\bibinfo {year} {2015})}\BibitemShut {NoStop}%
\bibitem [{\citenamefont {Challen}\ \emph {et~al.}(2019)\citenamefont
  {Challen}, \citenamefont {Denny}, \citenamefont {Pitt}, \citenamefont
  {Gompels}, \citenamefont {Edwards},\ and\ \citenamefont
  {Tsaneva-Atanasova}}]{challen_artificial_2019}%
  \BibitemOpen
  \bibfield  {author} {\bibinfo {author} {\bibfnamefont {R.}~\bibnamefont
  {Challen}}, \bibinfo {author} {\bibfnamefont {J.}~\bibnamefont {Denny}},
  \bibinfo {author} {\bibfnamefont {M.}~\bibnamefont {Pitt}}, \bibinfo {author}
  {\bibfnamefont {L.}~\bibnamefont {Gompels}}, \bibinfo {author} {\bibfnamefont
  {T.}~\bibnamefont {Edwards}}, \ and\ \bibinfo {author} {\bibfnamefont
  {K.}~\bibnamefont {Tsaneva-Atanasova}},\ }\href {\doibase
  10.1136/bmjqs-2018-008370} {\bibfield  {journal} {\bibinfo  {journal} {BMJ
  Quality \& Safety}\ }\textbf {\bibinfo {volume} {28}},\ \bibinfo {pages}
  {231} (\bibinfo {year} {2019})}\BibitemShut {NoStop}%
\bibitem [{\citenamefont {Courtland}(2018)}]{courtland_bias_2018}%
  \BibitemOpen
  \bibfield  {author} {\bibinfo {author} {\bibfnamefont {R.}~\bibnamefont
  {Courtland}},\ }\href {\doibase 10.1038/d41586-018-05469-3} {\bibfield
  {journal} {\bibinfo  {journal} {Nature}\ }\textbf {\bibinfo {volume} {558}},\
  \bibinfo {pages} {357} (\bibinfo {year} {2018})}\BibitemShut {NoStop}%
\bibitem [{\citenamefont {Wallach}\ and\ \citenamefont
  {Heifets}(2018)}]{wallach_most_2018}%
  \BibitemOpen
  \bibfield  {author} {\bibinfo {author} {\bibfnamefont {I.}~\bibnamefont
  {Wallach}}\ and\ \bibinfo {author} {\bibfnamefont {A.}~\bibnamefont
  {Heifets}},\ }\href {\doibase 10.1021/acs.jcim.7b00403} {\bibfield  {journal}
  {\bibinfo  {journal} {Journal of Chemical Information and Modeling}\ }\textbf
  {\bibinfo {volume} {58}},\ \bibinfo {pages} {916} (\bibinfo {year}
  {2018})}\BibitemShut {NoStop}%
\bibitem [{\citenamefont {McCloskey}\ \emph {et~al.}(2018)\citenamefont
  {McCloskey}, \citenamefont {Taly}, \citenamefont {Monti}, \citenamefont
  {Brenner},\ and\ \citenamefont {Colwell}}]{mccloskey_using_2018}%
  \BibitemOpen
  \bibfield  {author} {\bibinfo {author} {\bibfnamefont {K.}~\bibnamefont
  {McCloskey}}, \bibinfo {author} {\bibfnamefont {A.}~\bibnamefont {Taly}},
  \bibinfo {author} {\bibfnamefont {F.}~\bibnamefont {Monti}}, \bibinfo
  {author} {\bibfnamefont {M.~P.}\ \bibnamefont {Brenner}}, \ and\ \bibinfo
  {author} {\bibfnamefont {L.~J.}\ \bibnamefont {Colwell}},\ }\href
  {http://arxiv.org/abs/1811.11310} {\bibfield  {journal} {\bibinfo  {journal}
  {CoRR}\ }\textbf {\bibinfo {volume} {abs/1811.11310}} (\bibinfo {year}
  {2018})},\ \Eprint {http://arxiv.org/abs/1811.11310} {arXiv:1811.11310}
  \BibitemShut {NoStop}%
\bibitem [{\citenamefont {Stephen}\ and\ \citenamefont
  {Brodie}(2018)}]{stephen_brivaracetam:_2018}%
  \BibitemOpen
  \bibfield  {author} {\bibinfo {author} {\bibfnamefont {L.~J.}\ \bibnamefont
  {Stephen}}\ and\ \bibinfo {author} {\bibfnamefont {M.~J.}\ \bibnamefont
  {Brodie}},\ }\href {\doibase 10.1177/1756285617742081} {\bibfield  {journal}
  {\bibinfo  {journal} {Therapeutic Advances in Neurological Disorders}\
  }\textbf {\bibinfo {volume} {11}},\ \bibinfo {pages} {175628561774208}
  (\bibinfo {year} {2018})}\BibitemShut {NoStop}%
\bibitem [{\citenamefont {Amar}\ \emph {et~al.}(2019)\citenamefont {Amar},
  \citenamefont {Schweidtmann}, \citenamefont {Deutsch}, \citenamefont {Cao},\
  and\ \citenamefont {Lapkin}}]{amar_machine_2019}%
  \BibitemOpen
  \bibfield  {author} {\bibinfo {author} {\bibfnamefont {Y.}~\bibnamefont
  {Amar}}, \bibinfo {author} {\bibfnamefont {A.~M.}\ \bibnamefont
  {Schweidtmann}}, \bibinfo {author} {\bibfnamefont {P.}~\bibnamefont
  {Deutsch}}, \bibinfo {author} {\bibfnamefont {L.}~\bibnamefont {Cao}}, \ and\
  \bibinfo {author} {\bibfnamefont {A.}~\bibnamefont {Lapkin}},\ }\href
  {\doibase 10.1039/C9SC01844A} {\bibfield  {journal} {\bibinfo  {journal}
  {Chemical Science}\ }\textbf {\bibinfo {volume} {10}},\ \bibinfo {pages}
  {6697} (\bibinfo {year} {2019})}\BibitemShut {NoStop}%
\bibitem [{\citenamefont {Amar}(2018)}]{amar_accelerating_2018}%
  \BibitemOpen
  \bibfield  {author} {\bibinfo {author} {\bibfnamefont {Y.}~\bibnamefont
  {Amar}},\ }\emph {\bibinfo {title} {Accelerating process development of
  complex chemical reactions}},\ \href@noop {} {Ph.D. thesis} (\bibinfo {year}
  {2018})\BibitemShut {NoStop}%
\bibitem [{\citenamefont {Holbeck}\ \emph {et~al.}(2017)\citenamefont
  {Holbeck}, \citenamefont {Camalier}, \citenamefont {Crowell}, \citenamefont
  {Govindharajulu}, \citenamefont {Hollingshead}, \citenamefont {Anderson},
  \citenamefont {Polley}, \citenamefont {Rubinstein}, \citenamefont
  {Srivastava}, \citenamefont {Wilsker}, \citenamefont {Collins},\ and\
  \citenamefont {Doroshow}}]{holbeck_national_2017}%
  \BibitemOpen
  \bibfield  {author} {\bibinfo {author} {\bibfnamefont {S.~L.}\ \bibnamefont
  {Holbeck}}, \bibinfo {author} {\bibfnamefont {R.}~\bibnamefont {Camalier}},
  \bibinfo {author} {\bibfnamefont {J.~A.}\ \bibnamefont {Crowell}}, \bibinfo
  {author} {\bibfnamefont {J.~P.}\ \bibnamefont {Govindharajulu}}, \bibinfo
  {author} {\bibfnamefont {M.}~\bibnamefont {Hollingshead}}, \bibinfo {author}
  {\bibfnamefont {L.~W.}\ \bibnamefont {Anderson}}, \bibinfo {author}
  {\bibfnamefont {E.}~\bibnamefont {Polley}}, \bibinfo {author} {\bibfnamefont
  {L.}~\bibnamefont {Rubinstein}}, \bibinfo {author} {\bibfnamefont
  {A.}~\bibnamefont {Srivastava}}, \bibinfo {author} {\bibfnamefont
  {D.}~\bibnamefont {Wilsker}}, \bibinfo {author} {\bibfnamefont {J.~M.}\
  \bibnamefont {Collins}}, \ and\ \bibinfo {author} {\bibfnamefont {J.~H.}\
  \bibnamefont {Doroshow}},\ }\href {\doibase 10.1158/0008-5472.CAN-17-0489}
  {\bibfield  {journal} {\bibinfo  {journal} {Cancer Research}\ }\textbf
  {\bibinfo {volume} {77}},\ \bibinfo {pages} {3564} (\bibinfo {year}
  {2017})}\BibitemShut {NoStop}%
\bibitem [{\citenamefont {Chuang}\ and\ \citenamefont
  {Keiser}(2018)}]{note:chuang_comment_2018}%
  \BibitemOpen
  \bibfield  {author} {\bibinfo {author} {\bibfnamefont {K.~V.}\ \bibnamefont
  {Chuang}}\ and\ \bibinfo {author} {\bibfnamefont {M.~J.}\ \bibnamefont
  {Keiser}},\ }\href {\doibase 10.1126/science.aat8603} {\bibfield  {journal}
  {\bibinfo  {journal} {Science}\ }\textbf {\bibinfo {volume} {362}},\ \bibinfo
  {pages} {eaat8603. Note: Our own analysis indicates that the modelling
  approach pursued in Ahneman {\it et al.}'s original work {\it does} perform
  noticeably better than random, but that the validation routine employed
  drastically overstates the prediction performance due to combination bias.}
  (\bibinfo {year} {2018})}\BibitemShut {NoStop}%
\bibitem [{\citenamefont {Ghiringhelli}\ \emph {et~al.}(2015)\citenamefont
  {Ghiringhelli}, \citenamefont {Vybiral}, \citenamefont {Levchenko},
  \citenamefont {Draxl},\ and\ \citenamefont
  {Scheffler}}]{ghiringhelli_big_2015}%
  \BibitemOpen
  \bibfield  {author} {\bibinfo {author} {\bibfnamefont {L.~M.}\ \bibnamefont
  {Ghiringhelli}}, \bibinfo {author} {\bibfnamefont {J.}~\bibnamefont
  {Vybiral}}, \bibinfo {author} {\bibfnamefont {S.~V.}\ \bibnamefont
  {Levchenko}}, \bibinfo {author} {\bibfnamefont {C.}~\bibnamefont {Draxl}}, \
  and\ \bibinfo {author} {\bibfnamefont {M.}~\bibnamefont {Scheffler}},\ }\href
  {\doibase 10.1103/PhysRevLett.114.105503} {\bibfield  {journal} {\bibinfo
  {journal} {Physical Review Letters}\ }\textbf {\bibinfo {volume} {114}}
  (\bibinfo {year} {2015}),\ 10.1103/PhysRevLett.114.105503}\BibitemShut
  {NoStop}%
\bibitem [{\citenamefont {Ouyang}\ \emph {et~al.}(2018)\citenamefont {Ouyang},
  \citenamefont {Curtarolo}, \citenamefont {Ahmetcik}, \citenamefont
  {Scheffler},\ and\ \citenamefont {Ghiringhelli}}]{ouyang_sisso:_2018}%
  \BibitemOpen
  \bibfield  {author} {\bibinfo {author} {\bibfnamefont {R.}~\bibnamefont
  {Ouyang}}, \bibinfo {author} {\bibfnamefont {S.}~\bibnamefont {Curtarolo}},
  \bibinfo {author} {\bibfnamefont {E.}~\bibnamefont {Ahmetcik}}, \bibinfo
  {author} {\bibfnamefont {M.}~\bibnamefont {Scheffler}}, \ and\ \bibinfo
  {author} {\bibfnamefont {L.~M.}\ \bibnamefont {Ghiringhelli}},\ }\href
  {\doibase 10.1103/PhysRevMaterials.2.083802} {\bibfield  {journal} {\bibinfo
  {journal} {Physical Review Materials}\ }\textbf {\bibinfo {volume} {2}}
  (\bibinfo {year} {2018}),\ 10.1103/PhysRevMaterials.2.083802}\BibitemShut
  {NoStop}%
\bibitem [{\citenamefont {Coles}(2001)}]{coles_introduction_2001}%
  \BibitemOpen
  \bibfield  {author} {\bibinfo {author} {\bibfnamefont {S.}~\bibnamefont
  {Coles}},\ }\href {\doibase 10.1007/978-1-4471-3675-0} {\emph {\bibinfo
  {title} {An {Introduction} to {Statistical} {Modeling} of {Extreme}
  {Values}}}},\ Springer {Series} in {Statistics}\ (\bibinfo  {publisher}
  {Springer London},\ \bibinfo {address} {London},\ \bibinfo {year}
  {2001})\BibitemShut {NoStop}%
\bibitem [{\citenamefont {Fan}\ and\ \citenamefont {Lv}(2008)}]{fan_sure_2008}%
  \BibitemOpen
  \bibfield  {author} {\bibinfo {author} {\bibfnamefont {J.}~\bibnamefont
  {Fan}}\ and\ \bibinfo {author} {\bibfnamefont {J.}~\bibnamefont {Lv}},\
  }\href {\doibase 10.1111/j.1467-9868.2008.00674.x} {\bibfield  {journal}
  {\bibinfo  {journal} {Journal of the Royal Statistical Society: Series B
  (Statistical Methodology)}\ }\textbf {\bibinfo {volume} {70}},\ \bibinfo
  {pages} {849} (\bibinfo {year} {2008})}\BibitemShut {NoStop}%
\bibitem [{\citenamefont {Beker}\ \emph {et~al.}(2019)\citenamefont {Beker},
  \citenamefont {Gajewska}, \citenamefont {Badowski},\ and\ \citenamefont
  {Grzybowski}}]{beker_prediction_2019}%
  \BibitemOpen
  \bibfield  {author} {\bibinfo {author} {\bibfnamefont {W.}~\bibnamefont
  {Beker}}, \bibinfo {author} {\bibfnamefont {E.~P.}\ \bibnamefont {Gajewska}},
  \bibinfo {author} {\bibfnamefont {T.}~\bibnamefont {Badowski}}, \ and\
  \bibinfo {author} {\bibfnamefont {B.~A.}\ \bibnamefont {Grzybowski}},\ }\href
  {\doibase 10.1002/anie.201806920} {\bibfield  {journal} {\bibinfo  {journal}
  {Angewandte Chemie International Edition}\ }\textbf {\bibinfo {volume}
  {58}},\ \bibinfo {pages} {4515} (\bibinfo {year} {2019})}\BibitemShut
  {NoStop}%
\bibitem [{\citenamefont {Bart{\'o}k}\ \emph {et~al.}(2013)\citenamefont
  {Bart{\'o}k}, \citenamefont {Kondor},\ and\ \citenamefont
  {Cs{\'a}nyi}}]{bartok_representing_2013}%
  \BibitemOpen
  \bibfield  {author} {\bibinfo {author} {\bibfnamefont {A.~P.}\ \bibnamefont
  {Bart{\'o}k}}, \bibinfo {author} {\bibfnamefont {R.}~\bibnamefont {Kondor}},
  \ and\ \bibinfo {author} {\bibfnamefont {G.}~\bibnamefont {Cs{\'a}nyi}},\
  }\href {\doibase 10.1103/PhysRevB.87.184115} {\bibfield  {journal} {\bibinfo
  {journal} {Physical Review B}\ }\textbf {\bibinfo {volume} {87}} (\bibinfo
  {year} {2013}),\ 10.1103/PhysRevB.87.184115}\BibitemShut {NoStop}%
\bibitem [{\citenamefont {De}\ \emph {et~al.}(2016)\citenamefont {De},
  \citenamefont {Bart{\'o}k}, \citenamefont {Cs{\'a}nyi},\ and\ \citenamefont
  {Ceriotti}}]{de_comparing_2016}%
  \BibitemOpen
  \bibfield  {author} {\bibinfo {author} {\bibfnamefont {S.}~\bibnamefont
  {De}}, \bibinfo {author} {\bibfnamefont {A.~P.}\ \bibnamefont {Bart{\'o}k}},
  \bibinfo {author} {\bibfnamefont {G.}~\bibnamefont {Cs{\'a}nyi}}, \ and\
  \bibinfo {author} {\bibfnamefont {M.}~\bibnamefont {Ceriotti}},\ }\href
  {\doibase 10.1039/C6CP00415F} {\bibfield  {journal} {\bibinfo  {journal}
  {Physical Chemistry Chemical Physics}\ }\textbf {\bibinfo {volume} {18}},\
  \bibinfo {pages} {13754} (\bibinfo {year} {2016})}\BibitemShut {NoStop}%
\bibitem [{\citenamefont {Han}\ \emph {et~al.}(2017)\citenamefont {Han},
  \citenamefont {Wang}, \citenamefont {Gu}, \citenamefont {Dong},\ and\
  \citenamefont {Zhang}}]{han_asymmetric_2017}%
  \BibitemOpen
  \bibfield  {author} {\bibinfo {author} {\bibfnamefont {Z.}~\bibnamefont
  {Han}}, \bibinfo {author} {\bibfnamefont {R.}~\bibnamefont {Wang}}, \bibinfo
  {author} {\bibfnamefont {G.}~\bibnamefont {Gu}}, \bibinfo {author}
  {\bibfnamefont {X.-Q.}\ \bibnamefont {Dong}}, \ and\ \bibinfo {author}
  {\bibfnamefont {X.}~\bibnamefont {Zhang}},\ }\href {\doibase
  10.1039/C7CC01626C} {\bibfield  {journal} {\bibinfo  {journal} {Chemical
  Communications}\ }\textbf {\bibinfo {volume} {53}},\ \bibinfo {pages} {4226}
  (\bibinfo {year} {2017})}\BibitemShut {NoStop}%
\bibitem [{\citenamefont {Sidorov}\ \emph {et~al.}(2018)\citenamefont
  {Sidorov}, \citenamefont {Naulaerts}, \citenamefont {Ariey-Bonnet},
  \citenamefont {Pasquier},\ and\ \citenamefont
  {Ballester}}]{sidorov_predicting_2018}%
  \BibitemOpen
  \bibfield  {author} {\bibinfo {author} {\bibfnamefont {P.}~\bibnamefont
  {Sidorov}}, \bibinfo {author} {\bibfnamefont {S.}~\bibnamefont {Naulaerts}},
  \bibinfo {author} {\bibfnamefont {J.}~\bibnamefont {Ariey-Bonnet}}, \bibinfo
  {author} {\bibfnamefont {E.}~\bibnamefont {Pasquier}}, \ and\ \bibinfo
  {author} {\bibfnamefont {P.}~\bibnamefont {Ballester}},\ }\href {\doibase
  10.1101/504076} {\bibfield  {journal} {\bibinfo  {journal} {bioRxiv}\ }
  (\bibinfo {year} {2018}),\ 10.1101/504076}\BibitemShut {NoStop}%
\bibitem [{\citenamefont {Mason}\ \emph {et~al.}(2017)\citenamefont {Mason},
  \citenamefont {Stott}, \citenamefont {Ashenden}, \citenamefont {Weinstein},
  \citenamefont {Karakoc}, \citenamefont {Meral}, \citenamefont {Kuru},
  \citenamefont {Bender},\ and\ \citenamefont {Cokol}}]{mason_prediction_2017}%
  \BibitemOpen
  \bibfield  {author} {\bibinfo {author} {\bibfnamefont {D.~J.}\ \bibnamefont
  {Mason}}, \bibinfo {author} {\bibfnamefont {I.}~\bibnamefont {Stott}},
  \bibinfo {author} {\bibfnamefont {S.}~\bibnamefont {Ashenden}}, \bibinfo
  {author} {\bibfnamefont {Z.~B.}\ \bibnamefont {Weinstein}}, \bibinfo {author}
  {\bibfnamefont {I.}~\bibnamefont {Karakoc}}, \bibinfo {author} {\bibfnamefont
  {S.}~\bibnamefont {Meral}}, \bibinfo {author} {\bibfnamefont
  {N.}~\bibnamefont {Kuru}}, \bibinfo {author} {\bibfnamefont {A.}~\bibnamefont
  {Bender}}, \ and\ \bibinfo {author} {\bibfnamefont {M.}~\bibnamefont
  {Cokol}},\ }\href {\doibase 10.1021/acs.jmedchem.7b00204} {\bibfield
  {journal} {\bibinfo  {journal} {Journal of Medicinal Chemistry}\ }\textbf
  {\bibinfo {volume} {60}},\ \bibinfo {pages} {3902} (\bibinfo {year}
  {2017})}\BibitemShut {NoStop}%
\end{thebibliography}%

\newpage

\section*{Supporting Information}

\subsection{Yield deconvolution}

\subsubsection{Formalism}

In a multi-component reaction system, some molecular factors are expected to have a larger impact on reaction yield than others, in particular if these factors play a part in one of the rate-limiting steps. A useful data postprocessing step is therefore to decompose the yield function onto molecular terms associated (i) with the action of the compounds individually and (ii) the {\em inter}action (or synergy) between pairs of compounds. We use an effective equilibrium constant approach,
\inserteq{
\gamma \frac{c_\textrm{prod}}{c_\textrm{educt}} = \exp(-\beta \Delta G),
\label{eq:equilibrium}
} % end eq
where $c_\textrm{prod}$ and $c_\textrm{educt}$ are the concentrations of product and educt, respectively. $\Delta G$ is the effective Gibbs free energy change. We decompose this free energy onto molecular terms, and, to keep the expressions short, limit ourselves to combinations of two compounds $A$ and $B$:
\inserteq{
\Delta G(AB) = \Delta G_0 + \Delta G_1(A) + \Delta G_1(B) + \Delta G_2(AB) \nonumber
} % end eq
 Generalization to more compounds is readily achieved: We simply need to think of $A$ as the compound whose effect we try to single out (e.g., the substrate) and $B$ as a collective label for the other molecules of the combination.
 
Assuming that $\Delta G_1$ and $\Delta G_2$ are small compared to $\beta = 1/kT$, we can expand eq.~\ref{eq:equilibrium}, to obtain a decomposition of the yield function $y(AB) \propto \frac{c_\textrm{prod}}{c_\textrm{educt}}$:
\inserteq{
y(AB) = y_0 + \Delta_1(A) + \Delta_1(B) + \Delta_2(AB)
} % end eq
In a first step we now subtract from $y(AB)$ the yield that we would expect for a randomly chosen compound $A'$, paired up with the same molecule/combination $B$:
\inserteq{
\varepsilon_{1}(A | B) & = y(AB) - \frac{1}{N_A} \sum_{A'} y(A'B) \nonumber \\
& = \Delta_1(A) - \langle \Delta_1 \rangle_A + \Delta_2(A,B) - \langle \Delta_2(B) \rangle_A \nonumber.
} % end eq
We thus eliminated $\Delta_1(B)$ from the yield function, and will refer to this conditional yield term as an insertion effect: $\varepsilon_{1}(A|B)$ is the excess of the reaction outcome that should be attributed to the action of $A$, and that we expect to change radically as we replace $A \rightarrow A'$. Note, however, that $\varepsilon_{1}(A|B)$ still contains an interaction term $\Delta_2(A,B)$. We can eliminate this term by marginalizing $\varepsilon_{1}(A | B)$ over all tested $B$:
\inserteq{
\varepsilon_1(A)& = \frac{1}{N_B} \sum_{B'} \varepsilon_{1}(A | B') \nonumber \\
& = \Delta_1(A) - \langle \Delta_1 \rangle_A + \langle \Delta_2(A) \rangle_B - \langle \Delta_2 \rangle_{AB}.
} % end eq
With the first-order terms at hand, the two-body contributions ($\varepsilon_2$) read
\inserteq{
\varepsilon_2(AB) &= y(AB) - y_0 - \varepsilon_1(A) - \varepsilon_1(B)  \nonumber \\
& = \Delta_2(AB).
} % end eq

The above holds for a simple two-factor system. In general, however, the bimolecular $\varepsilon_2$ comprises {\it all} pair-wise contributions to the yield function -- i.e., for a combination $ABC$:
\inserteq{
 \varepsilon_2(ABC) = \Delta_2(AB) + \Delta_2(AC) + \Delta_2(BC)
} % end eq.
We can apply the same strategy as above for $y(AB)$ to this expression, in order to single out specific terms on the right-hand side. We define
\inserteq{
\varepsilon_2(AB|C) = \varepsilon_2(ABC) - \frac{1}{N_A N_B} \sum_{A'} \sum_{B'} \varepsilon_2(A'B'C)
} % end eq
Finally, we marginalize over all $C$ to obtain an expression which should reduce to $\Delta_2(AB)$, provided that $\langle \Delta_2(*) \rangle_C = 0$:
\inserteq{
\varepsilon_2(AB) &= \frac{1}{N_C} \sum_{C'} \varepsilon_2(AB|C') \nonumber \\
& = \Delta_2(AB) - \langle \Delta_2 \rangle_{AB} + \nonumber \\
 + \langle \Delta_2(A) \rangle_C & - \langle \Delta_2 \rangle_{AC} + \langle \Delta_2(B)\rangle_C - \langle \Delta_2 \rangle_{BC}.
} % end eq

\subsubsection{Buchwald-Hartwig amination}

\begin{figure}[t]
\centering
\includegraphics[width=0.85\linewidth]{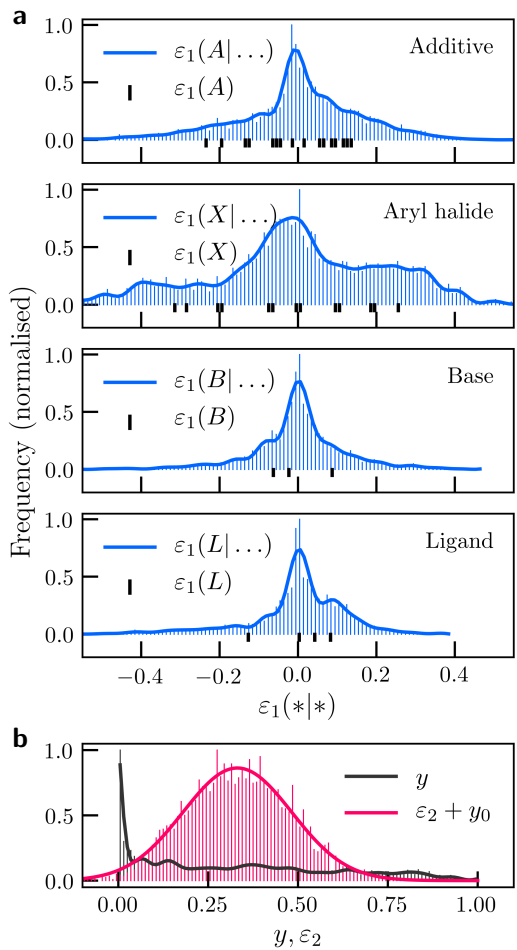}
\caption{ (a) Buchwald-Hartwig partial yields calculated from a molecular-term deconvolution of the reaction yield. Shown are distributions of the insertion effects $\pyield_1^{(i)}(.|\dots)$ of compound class $i$. The compound classes are, from top to bottom, the additive, aryl halide, base and ligand. The black dashes at the bottom of each graph indicate the unimolecular terms $\pyield_1^{(i)}$ associated with the individual compounds from each class. The distributions of total reaction yield $y$ (black histograms) and aggregated two-body terms (offset by the average yield $y_0$, see red histograms) are given in panel (b).}
\label{fig:sifig_yield_deconv_arx}
\end{figure}

Applied to the Buchwald-Hartwig system, the yield deconvolution gives insight into the relative importance of the four reagent/reactant classes regarding reaction outcome. The distributions of the insertion terms, shown as blue histograms in Fig.~\ref{fig:sifig_yield_deconv_arx}a, vary greatly in shape and width. Broadly speaking, the larger the width of the distribution, the larger the potential impact of this reagent type on reaction yield within the chemical space probed by the dataset. For the Buchwald-Hartwig reaction, the aryl halide (second from top) is most decisive, followed by the additive in second place. The effect of the base and ligand seems less important, but due to the small number of compounds tested for these classes (three and four, respectively) no definitive statements can be made regarding their impact.

In all cases, the distribution of insertion effects can be directly traced back to the unimolecular terms $\pyield_1^{(i)}(X)$, indicated in Fig.~\ref{fig:sifig_yield_deconv_arx}a as black dashes at the bottom of each histogram, with each dash corresponding to a single reagent. Note that the unimolecular terms are averages over a large number of samples (from 100 to 1000 and more) and hence have a smaller statistical error than the individual measurements. As an unfortunate side effect, however, the measurement errors as well as a ``coarse-graining'' error from the deconvolution procedure accumulate in the bimolecular terms, with the distribution of the total two-body contributions $\pyield_2$ assuming a distinctly Gaussian shape (see Fig.~\ref{fig:sifig_yield_deconv_arx}b, red curve). Nevertheless, almost the entire insertion effect is accounted for by unimolecular terms  -- revealing just how sparse these datasets are at heart: The combinatorial Buchwald-Hartwig data, despite testing thousands of unique combinations, are ultimately boiled down to 45 one-body terms (15 for the aryl halide, 23 for the additive, 3 for the ligand, 4 for the base) without a drastic loss of information.

\begin{figure*}[t]
\centering
\includegraphics[width=.95\linewidth]{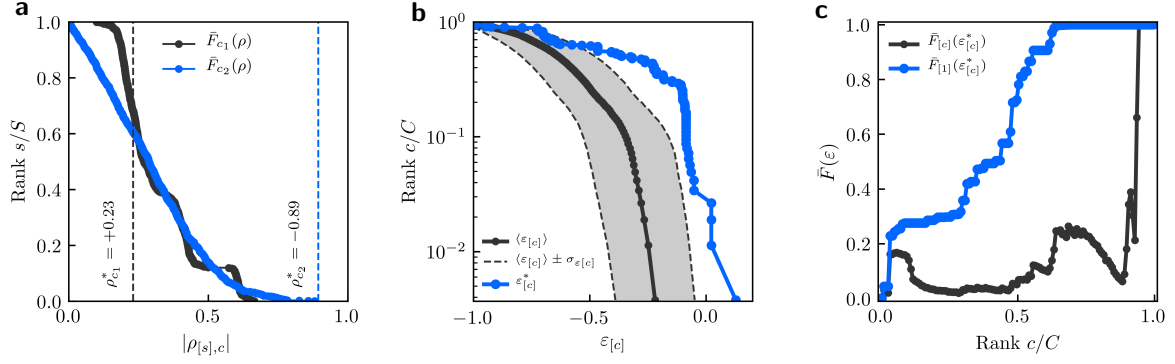}
\caption{Statistical feature-graph analysis in three steps: (a) Feature-target null correlation probabilities for two example nodes. (b) The global tail excess distribution as sampled across all random data feeds through the feature network (grey line and shaded area) compared to the ranked excesses measured for the physical system. (c) Assignment of feature p-values.}
\label{fig:sifig_stats}
\end{figure*}

\subsection{Network Filtering}

\subsubsection{Nonlinear covariance attribution}

Even simple nonlinear features can be difficult to interpret: The mathematical expression itself will not straightforwardly reveal which correlations between the variables determine the total covariance of the feature with the yield function. Some factors or terms appearing in a particular expression should be considered more significant, others less. Even the top-ranked function may include essentially irrelevant factors that can be safely omitted without compromising (or in fact, while improving) predictive power. In order to identify such factors, as well as to understand better which covariances drive the predictions, we attempt a decomposition of the covariance function $\rho = \rho(\varphi, y)$ onto body terms. Assuming that the feature $\varphi = \varphi(x_1, \dots, x_d)$ depends on $d$ base variables $x_1$ to $x_d$, this decomposition reads
\inserteq{
 \rho &= \rho_0 + \sum_i \rho_i + \sum_{i<j} \rho_{ij} + \sum_{i<j<k} \rho_{ijk} + \dots + \rho_{1 \dots d},
} % end eq
with indices $i,j,k \in \{ 1, \dots, d\}$, the covariance contributions $\rho_{i\dots j}$ (due to correlations among variables $x_i \dots x_j$) and a null covariance $\rho_0$. Here we use the signed covariance measure
\inserteq{
 \rho = \int f(x_1 \dots x_d) \varphi(x_1 \dots x_d) y(x_1 \dots x_d) \mathrm{d}^d\!\bm{x},
} % end eq
with the sample distribution $f$, the (z-scored) feature $\varphi$ and yield function $y$.

We now consider the effect on $\rho$ as we artificially decorrelate some ($n$) of the features from the ($d-n$) others when evaluating the expectation of the product $\varphi y$, by additionally integrating over a product of $n$ marginal (single-variable) distribution functions $f_{i}(x_i)$:
\inserteq{
 \bar{\rho}_{{i_1 \dots i_n}} &= \iint f(x_1 \dots x_d) \left( \prod_{r=1}^n f_{i_r}(x'_{i_r}) \right) \ \cdot \label{eq:rhobar} \\
 \cdot \ & \varphi(x_1 \dots x'_{i_1} \dots x'_{i_n} \dots x_d) y(x_1 \dots x_d) \mathrm{d}^d\!\bm{x} \ \mathrm{d}^n\!\bm{x}' \nonumber. 
} % end eq

We can express these partially randomized covariances $\bar{\rho}_{i_1\dots i_n}$ in terms of the body terms $\rho_{k\dots l}$ above -- seeing that $\rho_{k\dots l}$ is zero by definition if any of the $x_m \in \{x_k \dots x_l\}$ is among one of the $n$ base variables that were decorrelated from the remaining ones in the manner of eq.~\ref{eq:rhobar}. The connection between the $\bar{\rho}$'s and $\rho$'s therefore is
\inserteq{
 \bar{\rho}_{i_1 \dots i_n} = \rho_0 + \sum_{\substack{k \\ k \neq i_1 \dots i_n}} \!\! \rho_{k} + \sum_{\substack{k < l \\ k,l \neq i_1 \dots i_n}} \!\! \rho_{kl} + \dots
} % end eq
By evaluating $\bar{\rho}_{i_1 \dots i_n}$ for all unique combinations $\{ x_{i_1} \dots x_{i_n} \}$ of varying size $n$ ($0 \leq n \leq d$), we thus obtain a linear system of $2^d$ equations which can be solved uniquely for the $2^d$ partial covariances $\rho_0, \rho_i, \dots, \rho_{1 \dots d}$. The remaining issue is therefore how to compute the randomized covariances defined in eq.~\ref{eq:rhobar}. As the distribution functions $f$ and $f_{i_r}$ are only sparsely sampled, our approach to evaluating $\bar{\rho}_{i_1 \dots i_n}$ again rests on permutational sampling with $S$ permutation operator $n$-tuples $\bm{\pi}_s = \{ \pi_{s,i} \dots \pi_{s,j} \}$ with $\pi_{s,k} \in S_N$ ($n$ being the number of variables which we decorrelate, $N$ the number of samples):
\inserteq{
 \bar{\rho}_{i \dots j} = \frac{1}{S N} \sum_{s=1}^S \sum_{a = 1}^N &\varphi(x_{a,1} \dots x_{\pi_{s,i}(a),i} \dots \nonumber \\
 \dots & x_{\pi_{s,j}(a),j} \dots x_{a,d}) y_a.
} % end eq

\begin{figure*}[tb]%[tbhp]
\centering
\includegraphics[width=0.7\linewidth]{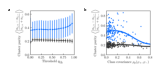}
\caption{ Mechanistic covariance clustering on synthetic data. (a) Average variation of achieved cluster purity with feature confidence threshold $q_\mathrm{th}$ (blue) compared to data with shuffled sample labels (grey). (b) Scatter plot of cluster purity vs class covariance between target functions $\varphi_+$ and $\varphi_-$ defining the two (hidden) classes. Each circle corresponds to one synthetic dataset. Grey circles indicate results with shuffled labels. Solid lines are regularized fits indicating how purity decays as class covariance increases (or, in other words, as the classes become mechanistically more similar). }
\label{fig:fig_covcluster}
\end{figure*}

\subsubsection{Mechanistic covariance clustering} The idea is to cluster samples based on their average covariance behaviour across a large set of nonlinear features. To this end, we partition the covariance measured between a target $y$ and function $\varphi_c$ onto contributions from individual samples: $\rho_c = \sum_{a=1}^N \Delta \rho_{ac}$. We address both the Pearson correlation $\rho_p$ and, for completeness, the signed AUC metric $\rho_\pm$: The sample contributions are
\inserteq{ 
  \Delta \rho_{p, ac} &= \left( \frac{ \varphi_{ac} - \langle \varphi_{c} \rangle) }{ \sigma_{\varphi_c} } \right) \left( \frac{  ( y_a - \langle y \rangle )}{ \sigma_y} \right), \\
  \Delta \rho_{\pm,[r(a)]c} &= \Delta t_{[r(a)]} (1 - f_{[r(a)]}) + \Delta f_{[r(a)]} t_{[r(a)]} - \frac{1}{N}. \nonumber
}
Here, $\langle \dots \rangle$ indicates an average taken over data samples $a \in \{ 1, \dots, N \}$. To understand the expression for the decomposition of $\rho_\pm$, consider the following recipe: We first sort the samples according to their feature value $\varphi_{a} \rightarrow \varphi_{[r(a)]}$, such that $r(a)$ is the rank corresponding to sample $a$; inversely, $a(r)$ is the sample $a$ corresponding to rank $r$. We subsequently track how the true positive rate $t_{[r]}$ and false-positive rate $f_{[r]}$ evolve as we increase the decision threshold $\varepsilon$ from rank 1 ($\varepsilon = \varphi_{[1]}$) to rank $N$ ($\varepsilon = \varphi_{[N]}$) -- in line with how the area under the receiver operating characteristic is usually determined. The true positive and false positive rates at rank $r$ are then defined recursively as $t_{[r]} = t_{[r-1]} + \Delta t_{[r]}$ and $f_{[r]} = f_{[r-1]} + \Delta f_{[r]}$, where $\Delta t_{[r]} = \delta(y_{a(r)} - y_+) / n_+$ and $\Delta f_{[r]} = \delta(y_{a(r)} - y_-) / n_-$ ($n_+$ and $n_-$ are the number of samples labeled as ``+'' and ``-'', respectively, $y_+$ and $y_-$ the associated class labels). Geometric inspection of the area under the $t(f)$ curve and transformation from AUC to $\rho_\pm = 2 \mathrm{AUC} -1 $ then leads to the expression for $\Delta \rho_{\pm,[r(a)]c}$ above.

Once the sample contributions are evaluated for all channels $c$ of interest, the distance measure follows from
\inserteq{
  d_{ab} = \frac{1}{C'} \sum_{c=1}^{C'} \frac{1}{2}\left( 1 - \frac{\Delta \rho_{ac} \Delta \rho_{bc}}{|\Delta \rho_{ac}| |\Delta \rho_{bc} |} \right).
}
We limit this sum to $C' \leq C$ channels, seeing that we may want to consider only a subset of the network nodes, specifically those with a confidence $q_c$ larger than some threshold $q_\mathrm{th}$.

A test on synthetic data shows that hierarchical clustering on the distance metric above indeed produces better-than-random enrichment across all cluster sizes. The synthetic data consists of 200 independently constructed datasets, each with 60 samples, and a 10-dimensional descriptor sampled from a uniform distribution over a finite domain, using a procedure equivalent to that described in the main text. A random split partitions the 60 samples onto two classes of size $n_+ = n_- = 30$, with the target function of class ``+'' chosen as $\varphi_+$, and of class ``-'' as $\varphi_-$ (both $\varphi_+$ and $\varphi_-$ are randomly selected nodes from the network). The descriptor alone therefore does not contain any information regarding which class a particular sample belongs to.

The outcome of the hierarchical clustering is evaluated using a purity 
\inserteq{ 
  m_i = \left| \frac{n_{i,+} - n_{i,-}}{n_{i,+} + n_{i,-}} \right|
}
evaluated for and then averaged over all clusters $i$ of the hierarchy, where $n_{i,+}$ and $n_{i,-}$ are the number of class members included in cluster $i$. Fig.~\ref{fig:fig_covcluster}a shows that the average purity slightly increases as the threshold $q_\mathrm{th}$ increases. Across the entire range of $q_\mathrm{th}$, the purity achieved by the clustering (blue) is significantly above the random baseline (black), estimated by shuffling the sample labels just before the purity is computed. The fluctuation of purities is, however, sizeable, partly because the clustering becomes more challenging as the covariance between the functions $\varphi_+$ and $\varphi_-$ increases: This is reflected in Fig.~\ref{fig:fig_covcluster}b, which correlates the achieved purity with the target covariance $\rho_p(\varphi_+, \varphi_-)$ for each dataset individually.

\begin{figure*}[t]%[tbhp]
\centering
\includegraphics[width=0.95\linewidth]{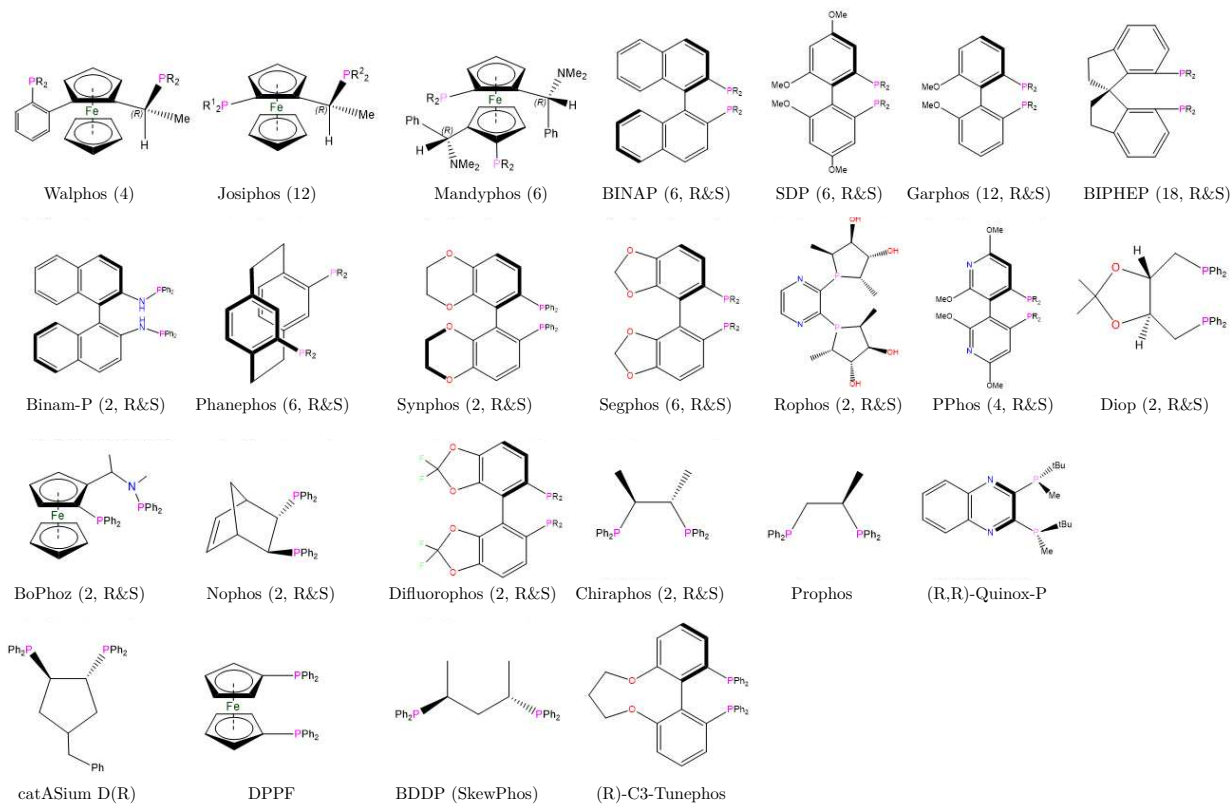}
\caption{ (a) Asymmetric hydrogenation: Chemical structures of ligands included in the experimental screen. }
\label{fig:fig_ligands_a}
\end{figure*}

\subsection{Physicochemical descriptors}

\subsubsection{Charge-transfer descriptors} We start from the potential energy surface of the molecule in the neutral, cationic, and anionic state -- denoted by $U_0(q)$, $U_+(q)$ and $U_-(q)$, respectively. The adiabatic ionization energy and electron affinity are given by
\inserteq{
\mathrm{IE} &= U_+(q_+) - U_0(q_0), \\
\mathrm{EA} &= U_-(q_-) - U_0(q_0).
} % end eq
where the nuclear coordinates $q_0$, $q_+$, $q_-$ are the equilibrium positions in the neutral, positively and negatively charged state. In addition to adiabatic, we also consider {\it vertical} IEs and EAs that do not incorporate nuclear relaxation:
\inserteq{ 
\mathrm{IE}_v &= U_+(q_0) - U_0(q_0), \\
\mathrm{IE}_v' &= U_+(q_+) - U_0(q_+).
}
Expressions for EA$_v$ and EA$_v'$ follow by analogy. We point out that DFT HOMO and LUMO energies are {\em not} suited to estimate these energies, even within Koopman's approximation. Instead they need to be determined from separate DFT calculations for each charge state.

The reorganization energies used in the physicochemical models are motivated by the semi-classical Marcus rate for charge transfer (i.e., non-adiabatic charge hopping) from a molecule $A$ to molecule $B$:
\inserteq{
k_{AB} = \frac{2\pi}{\hbar} \frac{|J_{AB}|^2}{\sqrt{4\pi\lambda_{AB} k_\textrm{B}T}} \exp \left[ - \frac{(\Delta U_{AB} - \lambda_{AB})^2}{4\lambda_{AB} k_{\textrm{B}}T} \right]
} % end eq
We estimate the reorganization energy $\lambda$ from four points on the potential energy surface of the molecules of the charge-transfer dimer:
\inserteq{
\lambda_{AB}^- = U_0^A(q_-^A) - U_0^A(q_0^A) + U_-^B(q_0^B) - U_-^B(q_-^B).
} % end eq
For hole transfer, an analogous expression is used, which substitutes $U_- \rightarrow U_+$ and $q_- \rightarrow q_+$. As the charge-transfer partners $A$ and $B$ are not known {\em a priori}, we use the unimolecular contributions $\lambda_A^{-0} = U_0^A(q_-^A) - U_0^A(q_0^A)$ (associated with discharging of $A$) and $\lambda_B^{0-} = U_-^B(q_0^B) - U_-^B(q_-^B)$ (charging of $B$) as descriptor elements. In order to reduce redundancy, the reorganization energies for hole and electron transfer, $\lambda_h$ and $\lambda_e$ used in the models are the respective averages for the discharging and charging process.

%In the Marcus-rate expression, $\Delta U_{AB}$ is the free energy difference between the initial and final state of the dimer. Assuming homogeneous stabilization of the molecular ions in solution, we incorporate the gas-phase unimolecular contributions to the numerator of the exponent (to be thought of as partial activation forces): These simplify to $g_{0-} = U_-(q_-) - U_0(q_-)$ and $g_{-0} = U_0(q_0) - U_-(q_0)$.

\subsubsection{Vibrational descriptors} The task is to automatically identify vibrational modes that are conserved across all molecules of a given class. The IR intensities (and in principle also the frequencies) can then be used as molecular descriptors. 

First, shared substructures of the compounds are identified. For the bisphosphine ligands, these are the PC$_3$ groups; for the additives, the isoxazole core; for the aryl halides, the aryl core. Normal modes, frequencies and intensities are calculated around the local energy minimum of a DFT ground state. We denote by $X_{Ai}$ the $n \times 3$ matrix of Cartesian displacements of a vibrational mode $i$ of molecule $A$. We define an $n \times n$ displacement correlation matrix $C_{Ai} = X_{Ai} X_{Ai}^t$ (note that this matrix may require symmetrization to match the underlying symmetries of the conserved substructures). $C_{Ai}$ satisfies the required permutational and rotational invariance, and can thus be used to compare two vibrational modes $i$ and $j$ of molecules $A$ and $B$, using a kernel function
\inserteq{ 
  k_{AB,ij} = \left(\frac{ C_{Ai} - \langle C_{Ai} \rangle } { \sigma_{C,Ai} } \right) : \left(\frac{ C_{Bj} - \langle C_{Bj} \rangle } { \sigma_{C,Bj} } \right),
}
where $\langle C \rangle$ denotes the component-wise average of matrix $C$, and $\sigma_C$ the corresponding standard deviation; ``:'' is the Frobenius inner product of the two matrices. The total kernel for modes $i$ and $j$ additionally takes into account a frequency term:
\inserteq{ 
  K_{AB,ij} = k_{AB,ij} \exp\left( - \frac{ (\nu_{Ai} - \nu_{Bj})^2}{2 \sigma_\nu^2 } \right),
}
with $\sigma_\nu = \unit[70]{cm}^{-1}$ to prevent the procedure from matching modes that are energetically remote. Given the complete matrix $K_{AB}$ of pairwise mode similarities $K_{AB,ij}$, the mode assignment is performed by optimizing an assignment matrix $P_{AB}$ such that $\bar{\kappa}_{AB} = P_{AB} : K_{AB}$ is maximized subject to the normalization constraints that $\sum_{i \in A} P_{AB,ij} = 1/N_B$, and $\sum_{j \in B} P_{AB,ij} = 1/N_A$ ($N_A$, $N_B$ are the number of vibrational modes of molecules $A$ and $B$). Mode $i$ of molecule $A$ is thus assigned to mode $j = \mathrm{argmax}_k (P_{AB,ik} K_{AB,ik})$ of $B$.

We finally identify conserved modes relative to a reference molecule $R$: A mode $r$ of $R$ is said to define a conserved mode if the average similarity $\bar{\kappa}_r = \frac{1}{N} \sum_{A, i} \kappa_{RA,ri}$ is larger than a threshold $\bar{\kappa}_\mathrm{th} = 0.8$, and all individual similarities satisfy $\bar{\kappa}_{RA,ri} \geq \kappa_\mathrm{th} = 0.3$. 

\begin{figure*}[t]%[tbhp]
\centering
\includegraphics[width=0.95\linewidth]{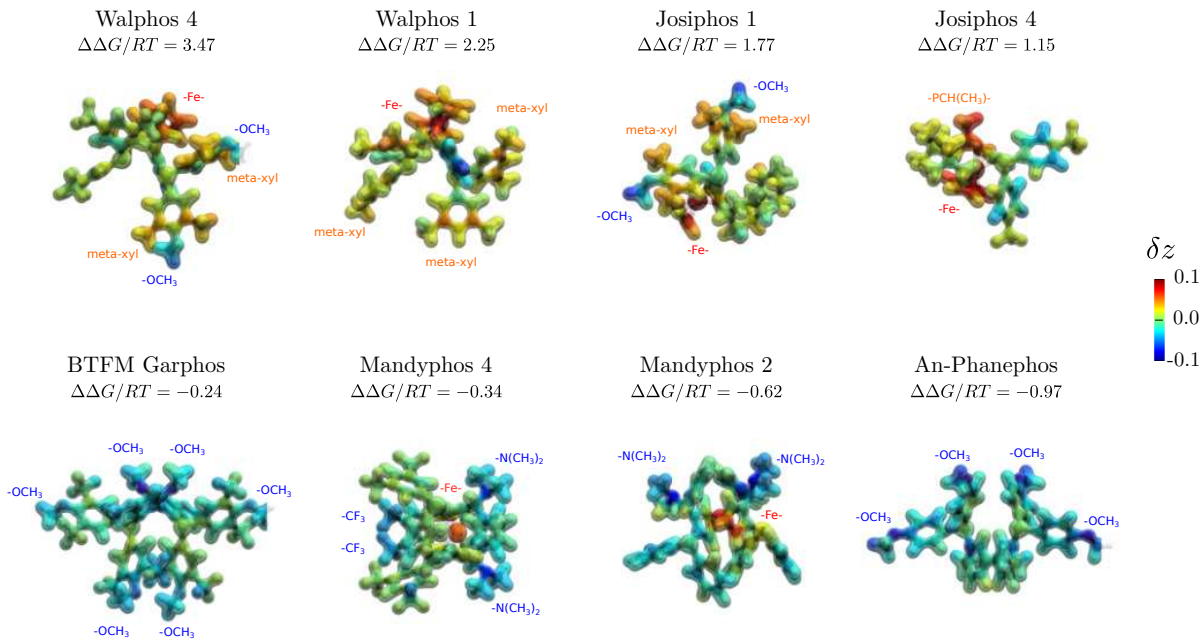}
\caption{ Atomic attribution for the classification of ligands into high- and low-$\Delta \Delta G^\ddagger$ compounds based on a molecular SOAP kernel. The top and bottom row each show four ligands associated with a large and small $\Delta \Delta G^\ddagger$, respectively, and a total conversion $y > 3\%$. An isosurface of the SOAP atomic density is coloured according to the value of atomic contributions $\delta z$ to an SVM decision function fitted to distinguish between high- and low-$\Delta \Delta G^\ddagger$ ligands (colour scale on the right). Red and blue moieties, for example, are strongly associated with a free-energy contribution that is larger and smaller, respectively, than the median $\Delta \Delta G^\ddagger$. }
\label{fig:fig_ligands_b}
\end{figure*}

\subsection{Asymmetric catalytic hydrogenation}

\subsubsection{Bisphosphine ligands} The chemical structures of the ligands included in the experimental screen are shown in Fig.~\ref{fig:fig_ligands_a}. Note that some of the ligands were tested in both their chiral R and S configuration. The difference in diastereomeric excess between chiral images was, however, generally small, with too few exceptions for us to be able to incorporate chirality into the model.

\begin{figure*}[t]%[tbhp]
\centering
\includegraphics[width=0.95\linewidth]{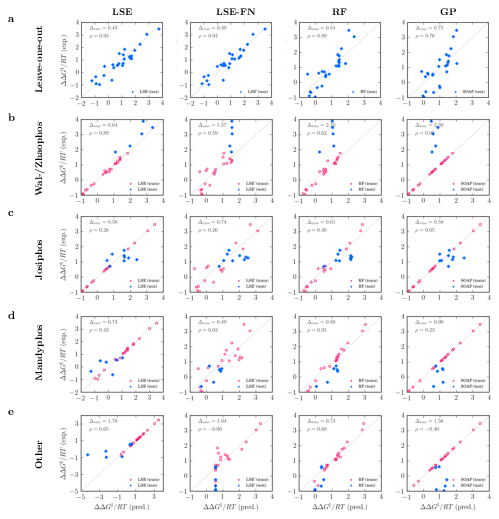}
\caption{ Leave-group-out (LGO) cross-validation on the asymmetric-catalysis dataset. The models tested are based on: linear least-squares ensembles (LSE), feature-network regressors (LSE-FN), random forests (RF) and Gaussian processes with a SOAP kernel (GP). Red open and blue closed circles are training and test predictions, respectively. }
\label{fig:fig_asymmcat_cv}
\end{figure*}

\subsubsection{Leave-group-out cross-validation} The ability of the models to generalize was tested using five-different cross-validation protocols (Fig.~\ref{fig:fig_asymmcat_cv}): Leave-one-out (LOO -- only one compound held out during the training), and leave-group out (LGO) for the Walphos/Zhaophos, Josiphos and Mandyphos scaffolds. The bottom row corresponds to withholding all compounds from the training set that are not part of any of these primary scaffold families.

\subsection{Hierarchical SOAP kernel}

The SOAP kernel for the Buchwald-Hartwig reactions and asymmetric hydrogenation consisted of two levels: The base-level kernel (comparing atomic environments $a$ and $b$) was $k_{ab} = (\bm{x}_a \cdot \bm{x}_b)^\xi$ with SOAP power spectra $\bm{x}_a$ and $\bm{x}_b$ and kernel exponent $\xi = 3$. The top-level kernel (comparing molecules $A$ and $B$) is 
\inserteq{ 
K^\eta_{AB} = \left(\sum_{a \in A} \sum_{b \in B} P_{ab} k_{ab}\right)^\eta.
}
Here, $\eta = 2$, and $P_{ab}$ is an assignment matrix optimized self-consistently so as to maximize $K_{AB}$, subject to the constraint that $\sum_a P_{ab} = 1/N_B$ and $\sum_b P_{ab} = 1/N_A$, where $N_A$ and $N_B$ are the numbers of atoms in structures $A$ and $B$ respectively.

For the modelling of growth-inhibitory synergy, a third kernel layer $\mathcal{K}$ is introduced as a similarity measure between molecular combinations $\mathcal{A} = (A,A')$ and $\mathcal{B} = (B,B')$:
\inserteq{ 
\mathcal{K}^\nu_{\mathcal{A}, \mathcal{B}} = \left(\sum_{A \in \mathcal{A}} \sum_{B \in \mathcal{B}} \mathcal{P}_{AB} K^\eta_{AB}\right)^\nu,
}
where $\nu = 4$. $\mathcal{P}$ is a molecular assignment matrix analogous to the atomic assignment matrix $P$ above. Regularization is achieved by adding a diagonal term $\delta \mathds{1}$ (with $\delta = 0.01$) to the top-level kernel. The kernel hyperparameters $\xi$, $\eta$, $\nu$ and $\delta$ as specified above are SOAP-specific and largely independent of the dataset at hand.

Built on top of an atomic kernel, the SOAP predictor function can be straightforwardly decomposed onto atomic contributions for insight. Unfortunately the (demonstrably) high learning capacity of SOAP-based frameworks does not respond well to sparse data. As a result, the structural attribution computed for the reaction systems studied here is of only limited value, because the predictor fails to capture key trends. Fig.~\ref{fig:fig_ligands_b} demonstrates this for the ligand effect on the asymmetric hydrogenation: Even though the molecular kernel correctly identifies relevant substituents, the associated weights are biased towards the Walphos and Josiphos class of ligands. This is most easily seen via the ferrocene cores, which are predicted to be primarily responsible for the high diastereoselectivity of those ligands -- contrary to the physicochemical modelling.

\end{document}